\documentclass[a4paper,11pt]{article}

\usepackage[a4paper,left=2.73cm,right=2.7cm,top=3cm,bottom=3.5cm]{geometry}

\usepackage{amsmath,amssymb,graphicx,hyperref,multirow,subcaption}
\usepackage[table]{xcolor}

\usepackage{parskip}
\def\d{\mathrm{d}}
\def\tr{\textrm{tr}\,}

\newcommand{\mt}[1]{\textrm{\tiny #1}}

\def\nc{N_\mt{c}}
\def\nf{N_\mt{f}}
\def\Rsym{\mt{R}}
\def\fla{\mt{f}}
\def\lef{\mt{L}}
\def\bar{\mt{b}}
\def\iso{\mt{I}}

\def\Mq{M_\mt{q}}

\newcommand{\be}{\begin{equation}}
\newcommand{\ee}{\end{equation}}

\newcommand{\bal}{\begin{aligned}}
\newcommand{\eal}{\end{aligned}}

\newcommand{\bea}{\begin{eqnarray}}
\newcommand{\beal}[1]{\begin{eqnarray}\label{#1}}
\newcommand{\eea}{\end{eqnarray}}

\numberwithin{equation}{section}

\newcommand{\fsub}{\mt{f}}
\newcommand{\sac}{\, , \qquad}
\newcommand{\eqn}[1]{Eq.~(\ref{#1})}
\newcommand{\Sec}[1]{Sec.~\ref{#1}}

\newcommand{\eqq}[1]{(\ref{#1})}
\newcommand{\fig}[1]{Fig.~\ref{#1}}

\newcommand{\mui}{\mu_\mt{I}}
\newcommand{\Nu}{n_\mt{u}}
\newcommand{\Nd}{n_\mt{d}}
\newcommand{\Nr}{n_\mt{R}}
\newcommand{\Ni}{n_\mt{I}}

\begin{document}

 \begin{titlepage}

\thispagestyle{empty}

\begin{flushright}
\hfill{ICCUB-19-011}\\
\hfill{HIP-2019-27/TH}
\end{flushright}

\vspace{40pt}  
	 
\begin{center}

{\huge \textbf{Spectrum of a Supersymmetric Color Superconductor}}

\vspace{30pt}
		
{\large \bf Ant\'on F. Faedo,$^{1}$   David Mateos,$^{1,\,2}$   \\ [1mm]
Christiana Pantelidou$^{3}$  and Javier Tarr\'\i o$^{4}$}

\vspace{25pt}

{\normalsize  $^{1}$ Departament de F\'\i sica Qu\'antica i Astrof\'\i sica and Institut de Ci\`encies del Cosmos (ICC),\\  Universitat de Barcelona, Mart\'\i\  i Franqu\`es 1, ES-08028, Barcelona, Spain.}\\
\vspace{15pt}
{ $^{2}$Instituci\'o Catalana de Recerca i Estudis Avan\c cats (ICREA), \\ Passeig Llu\'\i s Companys 23, ES-08010, Barcelona, Spain.}\\
\vspace{15pt}
{ $^{3}$Centre for Particle Theory and Department of Mathematical Sciences, Durham University, Durham, DH1 3LE, U.K.}\\
\vspace{15pt}
{ $^{4}$Department of Physics and Helsinki Institute of Physics, \\
P.O.Box 64, FIN-00014, University of Helsinki, Finland. 
}

\vspace{40pt}
				
\abstract{We have recently shown that the ground state of ${\cal N} = 4$, SU($\nc$) super Yang--Mills coupled to $\nf \ll \nc$ flavors, in the presence of non-zero isospin and R-symmetry charges, is a supersymmetric, superfluid, color superconductor. The holographic description consists of 
$\nf$ D7-brane probes in AdS$_5\times$S$^5$ with electric and instantonic fields on their worldvolume. These correspond to fundamental strings and D3-branes dissolved on the D7-branes, respectively. Here we use this description to determine the spectrum of mesonic excitations. As expected for a charged superfluid we find non-relativistic, massless Goldstone modes.  We also find extra ungapped modes that are not associated to the breaking of any global symmetries but to the supersymmetric nature of the ground state. If the quark mass is much smaller than the scale of spontaneous symmetry breaking   a pseudo-Goldstone boson is also present. We highlight some new features that appear only for $\nf> 2$. We show that, in the generic case of  unequal R-symmetry charges, the dissolved strings and D3-branes blow up into a D5-brane supertube stretched between the D7-branes.   
}

\end{center}

\end{titlepage}

\tableofcontents

\hrulefill
\vspace{10pt}

\section{Introduction}

The holographic duality \cite{Maldacena:1997re} makes the systematic   study of  a class of strongly coupled Quantum Field theories (QFT) possible (see e.g.~\cite{Erdmenger:2007cm,Ramallo:2013bua,Erdmenger:2018xqz} for some reviews). This framework, also referred to as AdS/CFT or gauge/gravity correspondence, arose in string theory by considering two complementary descriptions of the same system: a quantum gauge theory and a  theory of supergravity. The gravitational description of the system provides a geometric interpretation of the gauge theory dynamics by means of the so-called holographic dictionary, a prescription to translate between the QFT and gravity realisations \cite{Gubser:1998bc,Witten:1998qj}. The limit in which the gravitational side of the duality reduces to its classical limit --- which we will consider in this work--- takes the rank of the gauge group and the 't Hooft coupling to be very large, $\nc\to\infty$ and $\lambda\to\infty$.

In this paper we study a model at finite isospin density and vanishing temperature and baryon number, whose ground state simultaneously exhibits two of the properties that are believed to be present in some phases of QCD (see \cite{Alford:2007xm} for a review): spontaneous breaking of global symmetries and a Higgsing of the gauge symmetry, usually referred to as color superconductivity. Unlike our setup, in Nature systems that carry a large isospin density also carry a large (in fact, larger) baryon density, for example neutron stars. Another difference with realistic scenarios is that our ground state  is ${\cal N}=1$ supersymmetric. Nevertheless, we expect that our work can provide an interesting laboratory to understand the physics of cold quark matter. 

For example, we can develop  a first-principle understanding of strongly coupled matter in a regime where results from lattice field theory can be obtained and compared to. Lattice field theory encounters a serious limitation when considering configurations at finite baryon density, due to the infamous fermion sign problem \cite{deForcrand:2010ys}, but this difficulty is not present if instead of baryon density one considers a finite isospin density \cite{Son:2000xc}. Independently, supersymmetry can facilitate  a precise comparison between the strong-coupling limit described by holography and the weak-coupling regime accessible to perturbative field theory methods. 

Incidentally, to the best of our knowledge our system  is the first example of a supersymmetric color superconductor at finite density. Previous work on color superconductors in the non-supersymmetric holographic context includes \cite{Chen:2009kx,Basu:2011yg,Rozali:2012ry,BitaghsirFadafan:2018iqr}, and non-supersymmetric color superconducting ground states in supersymmetric theories have been considered in e.g.~\cite{Harnik:2003ke,Arai:2005pk,Rajput:2011zzc}.

The model we consider is very simple yet extremely rich: ${\cal N}=4$ SYM theory with SU($\nc$) gauge symmetry probed by $\nf\ll \nc$ ${\cal N}=2$ hypermultiplets transforming in the fundamental representation of the gauge group. Although this theory differs from Quantum Chromodynamics in many aspects, many insights about the behavior of strongly coupled quark matter were obtained by studying the holographic dual of the ${\cal N}=2$ theory in deconfined phases, be it static properties or far-from-equilibrium dynamics (see e.g.~\cite{CasalderreySolana:2011us} and references therein). For this reason, we use QCD nomenclature and refer to the adjoint degrees of freedom in ${\cal N}=4$ as gluons, or colors, and to the degrees of freedom in the fundamental representations as quarks, or flavors, despite the fact that these degrees of freedom come in supermultiplets.  We emphasise that at this point we do not aim at providing a model for real-world QCD but a theoretical laboratory in which first-principle calculations may lead to interesting insights \cite{Mateos:2011bs}.

The dual gravitational description  consists of a set of $\nf$ probe D7-branes (or flavor branes) in the AdS$_5\times$S$^5$  geometry sourced by a stack of $\nc$ coincident D3-branes (or color branes) \cite{Karch:2002sh}. The Higgsing of the gauge group is described by an instantonic configuration of the gauge field living on the worldvolume of the flavor branes \cite{Guralnik:2004ve,Guralnik:2004wq,Guralnik:2005jg}.

At zero density and $\nf>1$ this setup has a free parameter, the size of the instanton $\Lambda$, corresponding  holographically  to the scale of the spontaneous symmetry breaking. The presence of the instanton corresponds to D3-brane charge dissolved in the worldvolume of the D7-branes, which breaks the SU($\nc$) gauge group. The freedom to choose $\Lambda$ gives rise to a moduli of equivalent ground states, a Higgs branch parameterised by the vacuum expectation value of a scalar operator bilinear in the flavor hypermultiplets.  For $\nf=1$ there is no Higgs branch since the instanton collapses to zero size. This can be alleviated in the presence of a U(1) baryonic density \cite{Ammon:2012mu}, described by a radially-directed electric field on the worldvolume of the D7-brane \cite{Kobayashi:2006sb}: in the probe approximation the fields living on the flavor-brane feel an effective metric that hides the singularities of the solution behind a horizon, effectively regularising the solution.

We work with  $\nf \geq 2$ and introduce a non-Abelian (isospin) finite density. This was first considered, in a non-supersymmetric context, in \cite{Erdmenger:2007ap,Erdmenger:2007ja,Erdmenger:2008yj} for the D3/D7 setup and in \cite{Parnachev:2007bc,Aharony:2007uu} for the Sakai--Sugimoto model. For the solution to be supersymmetric the quarks are necessarily massive, with mass $\Mq$. We restrict our study to the critical case in which the isospin chemical potential has the same magnitude as the quark mass, $\mu_\iso=\Mq$. It is possible to define an Abelian U(1) symmetry under which the instanton is charged, such that one would expect the solution to collapse to zero size. The stabilization mechanism that allows the supersymmetric ground state to exist is a finite angular momentum in the compact directions of the D7-branes. The balance between the force due to the angular momentum and the tendency of the charged instanton to collapse determine uniquely the value of its size $\Lambda$. This size is therefore not a modulus  in our case.

The ground state of our system  breaks spontaneously various global symmetries, such that Goldstone modes are present in the spectrum. On top of this, the presence of a finite chemical potential breaks explicitly the Lorentz group, so we expect the dispersion relations of the Goldstone modes  to be non-relativistic. Indeed, we find that at low values of the momentum there are modes with a dispersion relation 
\be
\omega = \pm \frac{1}{2 \mu_\iso}\, k^2 \ .
\ee
When the relation $\Mq\ll \Lambda$ is satisfied there is a light excitation, a pseudo-Goldstone mode, with mass gap given by 
\be\label{eq.pseudogoldstonegap}
\omega_\mt{gap} = 2 \Mq \ .
\ee
This pseudo-Goldstone mode appears in the presence of the explicit symmetry-breaking scale $\Mq$. The approximate symmetry that is being broken is the scale invariance of the AdS$_5\times$S$^5$ background. This  becomes an exact symmetry when $\Mq=0$, which is then broken spontaneously by $\Lambda$. In that case the theory is Lorentz-invariant and one finds the relativistic dispersion relation $\omega=\pm k$ for the associated Goldstone boson.

In addition to the  above we find an additional set of ungapped modes that are not Goldstone modes. Their presence is due to the supersymmetric nature of the ground state, which implies the existence of exact moduli that are not associated to any broken global symmetries. 

The remaining excitations of our system have a finite life-time. These are characterised by quasifrequencies with a finite imaginary part lying in the lower complex-frequency plane (as necessary for stability). The complex quasifrequencies approach the real frequency axis when $\Lambda\ll\Mq$, and the imaginary parts vanish exactly when $\Lambda=0$. In that case there is no spontaneous symmetry breaking, and one recovers the mesonic spectrum of ${\cal N}=2$ SYM coupled to fundamental matter at finite isospin density \cite{Kruczenski:2003be,Erdmenger:2005bj}.

This paper is organised as follows. In  \Sec{sec.setup} we introduce a simple model capturing the string theory dynamics in supersymmetric configurations. The solution to this model with flavor group SU(2) was given in \cite{Faedo:2018fjw} and is thoroughly reviewed here, with an emphasis on the description of the spontaneous breaking of the local and global symmetries.
In  \Sec{sec.fluctuations} we proceed to study fluctuations around the supersymmetric solution. We pay  special attention to the description of the low-energy excitations of the system.
In most of the paper we focus on the case $\nf =2$. In  \Sec{higher} we discuss the generalisation to $\nf>2$. 
In \Sec{super} we show that, under generic circumstances, the strings and the D3-branes dissolved inside the D7-branes are ``blown up'' by the angular momentum to a D5-brane supertube . All other configurations can thus be viewed as collapsed supertubes.   
In \Sec{sec.conclusions} we present a summary of the physical implications of our results.
We complement the main text with several appendices to which we defer technical material.

\section{Setup\label{sec.setup}}

\subsection{Model}
\label{modelsec}
The holographic dual of ${\cal N}=4$ SU($\nc$) SYM theory in four dimensions is type IIB supergravity on the near-horizon geometry of a stack of $\nc$ coincident D3-branes, which source an AdS$_5\times$S$^5$ geometry with metric 
\be\label{eq.AdS5xS5metric}
\d s^2 = H^{-1/2} \, \eta_{\mu\nu} \, \d x^\mu \d x^\nu + H^{1/2} \left( \delta_{ij} \, \d y^i \d y^j + \delta_{\alpha\beta} \, \d z^\alpha \d z^\beta \right) \ ,
\ee
where $x^\mu$ (with $\mu=0,\cdots,3$) are the four gauge theory directions, and $y^i$ (with $i=1,\cdots,4$) and $z^\alpha$ (with $\alpha=1,2$) are Cartesian coordinates transverse to the D3-branes. Anticipating the setup we  split the six transverse coordinates as 
$
\mathbb{R}^6 \to \mathbb{R}^4 \times \mathbb{R}^2 \,.
$
 The function $H$ is a harmonic function in the six-dimensional space transverse to the D3-branes
\be\label{eq.harmonicfunction}
H = \frac{L^4}{\left( \delta_{ij} \, y^i y^j + \delta_{\alpha\beta} \, z^\alpha z^\beta \right)^2} \ , 
\ee
with $L$ the radius of AdS$_5$ and S$^5$, related to the string length $\ell_s$ and the 't Hooft coupling $\lambda$ through 
\be
L^4 = (2\pi\ell_s^2)^2 \frac{\lambda}{2\pi^2} \ .
\ee
In IIB supergravity the geometry \eqq{eq.AdS5xS5metric} is supported by a Ramond-Ramond (RR) four-form potential and a constant dilaton
\be
C_4 = H^{-1} \d x^{1,3} \ , \qquad e^\Phi= 1 \ ,
\ee
with $\d x^{1,3}$ the volume form of the Minkowski spacetime. The RR four-form potential determines the type IIB self-dual five-form as
\be
F_5=(1+ \star) \d C_4 \,,
\ee
 where $\star$ is the Hodge-star operator in ten dimensions. The flux of the RR five-form through the S$^5$ gives the number of D3-branes
\be\label{eq.F5flux}
\frac{1}{(2\pi\ell_s)^4}\int_{S^5} F_5 = \nc \ .
\ee

The ${\cal N}=4$  SYM theory does not contain any degrees of freedom in the fundamental representation of the gauge group,  but these can be included by adding $\nf$ D7-branes on the gravity side, oriented along the $x^\mu$- and  $y^i$-directions, as indicated in Table~\ref{tab.intersection}.
\begin{table}[t!]
 \centering
  \begin{tabular}{c c c cc c c c c c}
   {} & $\mathbb{R}^{1,3}$ & \cellcolor{green!15} $\mathbb{R}^4$ & \cellcolor{green!15}$z^1$ & \cellcolor{green!15} $z^2$ & & \multicolumn{2}{c}{\cellcolor{green!15} SO(6)$_\Rsym$} & & \\
   D3 & $\times$ & \cellcolor{blue!15} $\cdot$ & \cellcolor{yellow!15} $\cdot$ & \cellcolor{yellow!15} $\cdot$ & & \cellcolor{blue!15} SO(4) & \cellcolor{yellow!15} SO(2) & & \\ 
   D7 & $\times$ & \cellcolor{blue!15} $\times$ & \cellcolor{red!15} $+$ &  $\,\cdot$ & & \cellcolor{red!15} \cellcolor{blue!15} & & & \cellcolor{red!15} \\
   D7 & $\times$ & \cellcolor{blue!15} $\times$ & \cellcolor{red!15} $-$ & $\,\cdot$ & & \multirow{-2}{*}{\cellcolor{blue!15} SU(2)$_\lef \times$ SU(2)$_\Rsym$} & & & \multirow{-2}{*}{\cellcolor{red!15} U(1)$_\iso \times$U(1)$_\bar$} 
  \end{tabular}
 \caption{\label{tab.intersection} \small Diagram of the intersections and the explicit pattern of symmetry breaking. In our solution the two D7-branes bend in opposite directions in the $z^1$ direction, breaking SO(2).}
\end{table}
On the gauge theory side the new degrees of freedom are $\nf$ hypermultiplets in the fundamental representation, to which we will loosely refer as flavor or quarks despite the fact that they include both fermions and scalars. Coupling these fields to the original 
${\cal N}=4$  SYM theory explicitly breaks supersymmetry down to ${\cal N}=2$. We will therefore refer to the resulting theory simply as ``the  ${\cal N}=2$ theory''. 

The inclusion of the quarks in the model also breaks the R-symmetry group explicitly as 
$
\mbox{SO(6)}_\Rsym \to \mbox{SO(4)}\times \mbox{SO(2)} \,.
$
On the gravity side the breaking is due to the splitting of $\mathbb{R}^6$ induced by the orientation of the D7-branes. The remaining SO(4) and SO(2) factors act as rotations in the $y^i$- and \mbox{$z^\alpha$-directions}, respectively. For our purposes it will be convenient to view the rotation group acting on the $y^i$-directions as  
$
\mbox{SO(4)} =  \mbox{SU(2)}_\lef \times \mbox{SU(2)}_\Rsym  \,,
$
where SU(2)$_\Rsym$ is the R-charge symmetry group of the ${\cal N}=2$ theory and SU(2)$_\lef$ is a global symmetry that does not act on the ${\cal N}=2$ supercharges. On the gravity side this is made explicit by writing the metric in the $y^i$-space in spherical coordinates as
\be\label{eq.sphericalcoordinates}
\delta_{ij} \, \d y^i \d y^j = \d r^2 + r^2 \, \delta_{ab} \, \omega^a \omega^b \ ,
\ee
with $\omega^a$ ($a=1,2,3$) the left-invariant one-forms on S$^3$ satisfying
\be
\d \omega^a + \epsilon^{abc} \omega^b \wedge \omega^c = 0 \ .
\ee
As indicated by their name, the $\omega^a$ are invariant under SU(2)$_\lef$, and they transform as a triplet under SU(2)$_\Rsym$. 

In the regime $\nf\ll \nc$ the flavor degrees of freedom can be treated as probes of the color-dominated dynamics. On the supergravity side this means that the backreaction of the D7-branes on the geometry \eqref{eq.AdS5xS5metric} can be ignored \cite{Karch:2002sh}. The embedding of the D7-branes will be specified by their positions in the $z^\alpha$-plane. If the branes are all coincident and located at the origin of the plane then the SO(2) symmetry is preserved. In addition,  the flavor symmetry of the gauge theory, which is realised as the non-Abelian gauge symmetry on the worldvolume of the D7-branes, is  
\be
\mbox{{U($\nf$)$ = $SU($\nf$)$ \times $U(1)$_\bar$}}\,.
\ee
The Abelian factor on the right-hand side can be thought of as the symmetry associated to baryon charge, which will always be unbroken in this paper. Except in \Sec{higher}, in the rest of the paper we will set the number of flavors to 
\be
\nf=2 \,.
\ee
In generic cases, including those of interest here, the branes will have non-trivial profiles in the $z^\alpha$-plane. Under these circumstances SO(2) is broken to at most a discrete subgroup, and the flavor symmetry is broken to an Abelian subgroup:
\be
\label{implement}
\mbox{SU(2)}_\fla  \to \mbox{U(1)}_\iso \,.
\ee
We will refer to the charge under U(1)$_\iso$ as isospin charge. Recall that on the D7-branes the symmetries in \eqq{implement} are gauge symmetries. In the case under consideration of two D7-branes that bend in a generic way in the $z^\alpha$-plane, each D7-brane carries a U(1) gauge symmetry on its worldvolume. The surviving U(1)$_\bar$ and U(1)$\iso$ symmetries are  then simply the diagonal and the off-diagonal combinations of the two U(1)'s on the branes, respectively.

\subsection{Action}

The action  describing the dynamics of $\nf=2$ D7-branes embedded in AdS$_5 \times$ S$^5$ is given by two pieces, 
\be\label{eq.originalaction}
S_\mt{D7s} = S_\mt{DBI} + S_\mt{WZ} \ ,
\ee
where the non-Abelian Dirac--Born--Infeld (DBI) and Wess--Zumino actions (WZ) adapted to our system read \cite{Tseytlin:1997csa,Myers:1999ps} 
\be
\label{wvaction}
\bal
S_\mt{DBI} & = - T_7 \int \d^4 x\, \d^4 y \, e^{-\Phi} \, \mathrm{Str} \left[ \sqrt{-\det \left( g_{MN} + {\mathcal P}[H]^\frac{1}{2} \delta_{\alpha\beta}\, D_{(M}Z^\alpha D_{N)}Z^\beta +  F_{MN}  \right) }\,V(Z) \right]\ , \\
S_\mt{WZ} & = \frac{T_7}{2} \, \int \mathrm{Str} \left( C_4 \wedge F \wedge F \right) \,.
\eal
\ee
$T_7$ is the tension of a D7-brane and, unless otherwise indicated, in this paper we set $2\pi \ell_s^2 = 1$, so all quantities are effectively dimensionless. The indices $M, N=0, \cdots, 7$ are worldvolume indices on the D7-branes and collectively denote the $x^\mu$ and $y^i$ directions. The non-Abelian gauge field $A$ takes values in the Lie algebra of SU(2)$_\fla$. Thus, suppressing spacetime indices, we can write  
$
A=A_{\hat a}\,\sigma^{\hat a} \,,
$
with $\sigma^{\hat a}$ ($\hat a=1,2,3$) the Pauli matrices. Note that in this paper the component of the gauge field along the generator of the baryonic U(1)$_\bar$ symmetry is assumed to vanish. In other words, we will work at non-zero isospin density but zero baryon density.  $A$ enters the action through the field strength 
\be
F=\d A +i\, A \wedge A
\ee
and the covariant derivatives 
\be
D Z^\alpha=\d Z^\alpha+ i\, [A,Z^\alpha]
\ee
for the scalars $Z^\alpha$. These are also SU(2)$_\fla$-valued,
$
Z^\alpha = Z^{\alpha}_{\hat a} \,\sigma^{\hat a} \,,
$
and describe the (in general non-commuting)  positions of the branes in the $z^\alpha$-directions. The non-Abelian nature of the action leads to the presence of the potential term \cite{Myers:1999ps}
\be\label{eq.Zpotential}
V(Z) = \sqrt{ 1 -  \,{\mathcal P}[H] \, \left[Z^1,Z^2 \right]^2} \ .
\ee
The symmetrised-trace over the flavor group, $\mathrm{Str}$, allows one to treat the non-Abelian structures as effectively commuting in the action \cite{Tseytlin:1997csa,Myers:1999ps}.  The metric $g_{MN}$ in the action is given by 
\be\label{wvmetric}
\d s^2 = {\mathcal P}[H]^{-1/2} \, \eta_{\mu\nu} \, \d x^\mu \d x^\nu + {\mathcal P}[H]^{1/2} \,\delta_{ij} \, \d y^i \d y^j  \,,
\ee
with 
\be
{\mathcal P}[H] = \frac{L^4}{\left( \delta_{ij} \, y^i y^j +  \delta_{\alpha \beta}\,  Z^\alpha Z^\beta \right)^2} 
\ee
the pull-back of the harmonic function \eqref{eq.harmonicfunction} to the worldvolume.  

As emphasised by the authors of \cite{Tseytlin:1997csa,Myers:1999ps}  themselves, the action \eqq{wvaction} is known to be incomplete beyond $\mathcal{O}(\ell_s^4)$. However, it  seems to capture the exact physics for supersymmetric configurations \cite{Hashimoto:1997px,Bak:1998xp}. Moreover, for such configurations the equations of motion obtained from  \eqq{wvaction} reduce to those  obtained from its lowest-order approximation,  namely from the  super-Yang--Mills--Higgs-like (SYMH) action 
\be\label{eq.ourmodel}
\bal
S & = - \frac{T_7}{2} \int\mathrm{Str} \Big( F \wedge * F  - {\mathcal P}[H]^{-1} \d x^{1,3} \wedge F \wedge F \Big) \\
& \quad - \frac{T_7}{2} \int\mathrm{Str}\Big( {\mathcal P}[H]^{1/2} \delta_{\alpha\beta}\, D Z^\alpha \wedge * D Z^\beta - {\mathcal P}[H]\, \left[Z^1,Z^2\right]^2 *1 \Big) \,,
\eal\ee
where $*$ is the eight-dimensional Hodge-star dual with respect to the metric \eqq{wvmetric} and the potential for the $Z^\alpha$ scalars comes from expanding \eqref{eq.Zpotential}.  In Appendix \ref{app.naDBI2HYM} we  prove that, indeed, the equations of motion that follow from  \eqq{wvaction} and \eqq{eq.ourmodel} coincide with one another when evaluated on the type of supersymmetric configurations that we will consider. We will therefore work with \eqq{eq.ourmodel} instead of \eqq{wvaction}.

\subsection{Ansatz}
We will seek solutions that are stationary and homogeneous along the gauge theory spatial directions, thus effectively reducing the dynamics to   4+1 dimensions in the $\{t, y^i\}$ directions. Moreover, we will be interested in configurations that are a direct import of the five-dimensional dyonic instanton of SYMH theory in flat space \cite{Lambert:1999ua}. As such, only one of the two scalar fields will be excited
\be
\label{obey}
Z^1 \equiv Z(y) \ , \qquad Z^2 = 0 \ .
\ee
This explicitly breaks the SO(2) symmetry of the action \eqq{eq.ourmodel} associated to rotating the $Z^\alpha$ into one another. For configurations obeying  \eqq{obey} the potential for the $Z^\alpha$ scalars in \eqq{eq.ourmodel} vanishes identically. Furthermore, as in [37], we will take
\be\label{eq.BPS}
A_t(y) = Z(y) \sac F_{ij}(y) =  \frac{1}{2} \epsilon_{ijkl} F^{kl}(y)   \,.
\ee
The first condition in \eqref{eq.BPS} ensures that the non-Abelian electric field equals the covariant derivative of the scalar 
\be
\label{deri}
E_i=D_iZ \,. 
\ee
The second equation is the usual  self-duality condition of the magnetic part of the field strength, which gives rise to instantonic configurations. The first condition together with Gauss' law, $DE=0$, implies that the scalar field obeys a covariant Laplace equation  in the background of an instanton in $\mathbb{R}^4$: 
\be
 \delta^{ij} D_i D_j Z(y) = 0 \,.
\ee
Solutions of these equations preserve ${\cal N}=1$ supersymmetry. This is the key property behind the exact nature of these configurations. In the particular case $Z=0$, Refs.~\cite{Guralnik:2004ve} and \cite{Arean:2007nh} showed that corrections at leading order and to all orders in 
$\ell_s$, respectively, vanish identically. Ref.~\cite{Zamaklar:2000tc} showed that configurations with possibly non-zero $Z$ are solutions of the action \eqq{wvaction} in flat space. In Appendix \ref{app.naDBI2HYM} we show that these remain solutions of both \eqq{wvaction} and \eqq{eq.ourmodel} even in the case in which the background is AdS$_5\times$S$^5$.

In most of this paper we will focus on solutions that preserve the SU(2)$_\lef$ global symmetry. In this case all the  angular dependence is captured by the left-invariant one-forms $\omega^a$ introduced in \eqref{eq.sphericalcoordinates}. Under these circumstances the set of solutions that we will be interested in may be described with the following ansatz for the different fields:
\be
\label{eq.ansatz}
Z = \phi(r)\, \sigma^3 \ , \qquad A = a_t(r) \, \d t \otimes\sigma^3 + a(r) \, \delta_{a\, {\hat a}}\, w^a \otimes \sigma^{\hat a} \ .
\ee
Note that, in expressions like the one above, we will distinguish between indices $a=1,2,3$ that are acted upon by SU(2)$_\Rsym$ and indices $\hat a=1,2,3$ that are acted upon by SU(2)$_\fla$. 
Splitting the field strength into purely electric and purely magnetic components, 
\be
F = \d t \wedge E + F_\mt{mag} \,,
\ee
this ansatz leads to the following result: 
\be
\label{eq.electricmagneticfields}
\bal
E & = - \phi' \, \d r \otimes \sigma^3 + 2 \, \phi \, a \, \delta_{a \, \hat a} \, \epsilon^{a b 3} \, \omega^b \otimes \sigma^{\hat a} \ , \\[2mm]
F_\mt{mag} & = a'  \delta_{a\, \hat a}\, \d r \wedge \omega^a \otimes \sigma^{\hat a} -  \, a\, (1+a) \, \delta_{a \, \hat a} \, \epsilon^{abc} \, \omega^b \wedge \omega^c \otimes \sigma^{\hat a} \,,
\eal
\ee
where $'$ denotes differentiation with respect to $r$. 

We have aligned the scalar field, and by virtue of  \eqq{eq.BPS} also the time component of the gauge field, with the third generator of 
SU(2)$_\fla$. As anticipated above, solutions with non-zero $\phi(r)$ will break this symmetry explicitly as described in \eqq{implement}. The authors of \cite{Lambert:1999ua} used the preserved U(1)$_\iso$ to define the electric charge of the instanton as 
\be
\label{QQQ}
Q \equiv \lim_{r\to\infty} \frac{1}{\Mq} \int_{S^3} r^3 \, \tr\! \left( Z E_r \right) \propto  \Lambda^2 \Mq \ ,
\ee
where the integral is taken over the three-sphere of radius $r$ in Eqn.~\eqref{eq.sphericalcoordinates} and $\Lambda$ and $\Mq$ are constants of integration to be introduced shortly.

The second term in $A$, proportional to $a(r)$,   gives rise to the purely magnetic part of the field strength. In fact, a non-trivial $a(r)$ has two important consequences. First, it implies that the D7-branes carry an instanton number given by 
\be\label{eq.instantonnumber}
k = \frac{1}{8\pi^2} \int \tr F_\mt{mag}\wedge F_\mt{mag} = 
- \Big[ 3 a^2 + 2 a^3 \Big]^{r=\infty}_{r=0} \,.
\ee
In turn, this results in the partial breaking of the color symmetry in the gauge theory. Second, a non-trivial  $a(r)$  locks the triplet of SU(2)$_\Rsym$ one-forms $\omega^a$ to the triplet of SU(2)$_\fla$ generators $\sigma^{\hat a}$, thus also breaking some global symmetries. Both of these breakings will be explained in detail in \Sec{sec.symmetrypatterns}.

Note that the expressions \eqq{QQQ} and \eqq{eq.instantonnumber} are appropriate when we view the above construction as a dyonic instanton in flat $\mathbb{R}^{1,4}$. However, we will see below that these same expressions arise in the context of the eight-dimensional theory on the D7-branes in a curved  AdS$_5\times$S$^5$ background.

\subsection{Solution}

The BPS equations \eqref{eq.BPS} for our ansatz \eqq{eq.ansatz} become
\be
a_t=\phi \sac
\phi'' + \frac{3}{r} \phi' - \frac{8 \, a^2}{r^2} \phi = 0 \sac
 a' + \frac{2}{r} a(1+a) = 0 \ ,
\ee
whose  solution of interest to us is
\begin{subequations}
\label{eq.solution}
\begin{align}
a_t(r) = \phi(r) &= \Mq \, \frac{r^2}{r^2 + \Lambda^2} \,, \\[2mm]
a(r) &= - \frac{\Lambda^2}{r^2 + \Lambda^2} \,,
\label{ar}
\end{align}
\end{subequations}
with $\Mq$ and $\Lambda$ constants of integration. This solution has instanton number $k=1$.

\begin{figure}[t]
\begin{center}
\includegraphics[width=.6\textwidth]{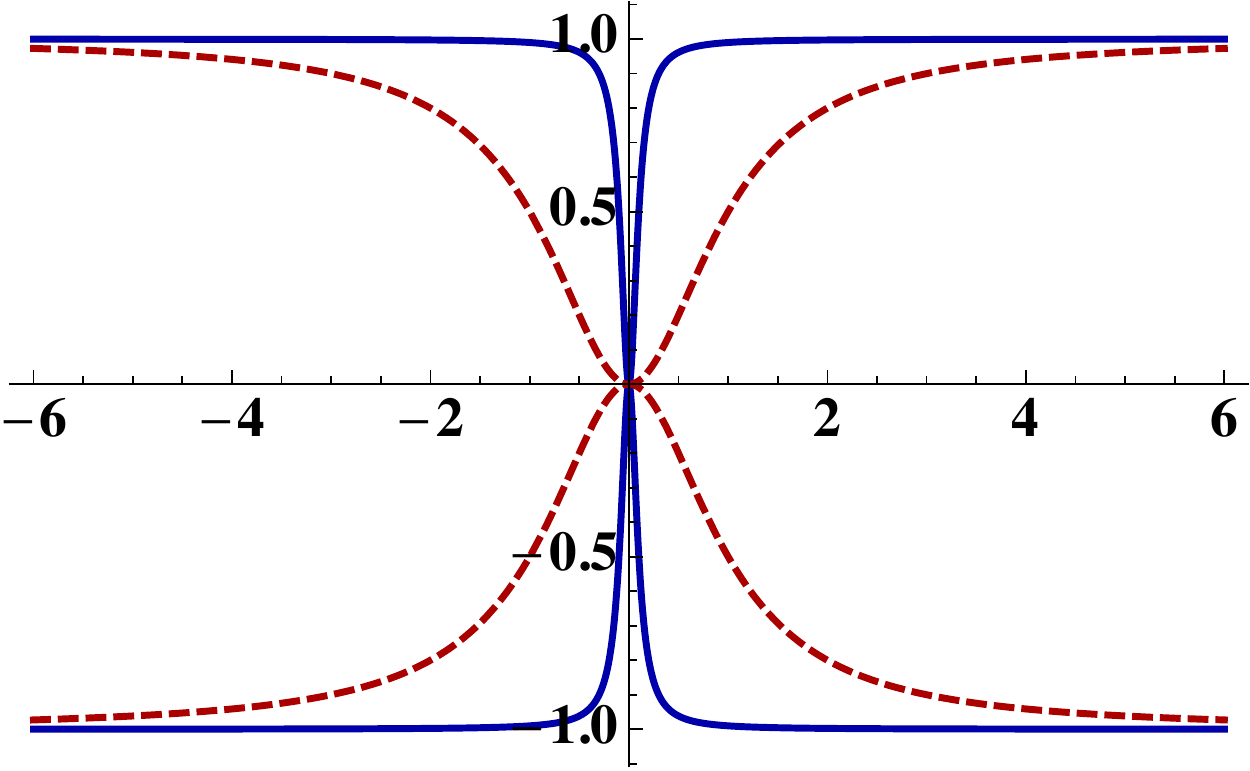}
\put(5,80){\Large$r/\Mq$}
\put(-155,173){\Large $\pm\phi/\Mq$}
\end{center}
\vspace{-1mm}
\caption{\small Embedding profile of the D7-branes in the $z^1$-direction for $\Lambda=\Mq$ (dashed, red curve) and $\Lambda=\Mq/10$ (continuous, blue curve). We have extended the range of the radial coordinate to negative values to represent a section of the solid of revolution in the $y^i$ directions.\label{fig.profile}}
\end{figure}

The physical interpretation of the solution is as follows. The fact that $Z(r)$ is proportional to $\sigma^3$, which is diagonal with entries $\pm 1$,  means that the branes bend in opposite directions along the \mbox{$z^1$-axis} with otherwise identical profiles, as shown in \fig{fig.profile}. 
The gauge theory operator dual to the scalar field has $\Delta=3$ and takes the schematic form \cite{Karch:2002sh}
\be
\label{Zdual}
Z = \phi\, \sigma^3 \,\,\,\,\,\, \leftrightarrow \,\,\,\,\,\, 
{\cal O} \, \sim \,\overline \Psi \sigma^3 \Psi  \, \sim \, \overline u u -  \overline d d   \,,
\ee
where $\Psi=(u,d)^T$ and $u, d$ are the  fermionic fields corresponding to the two quark flavors. In this expression we omitted squark contributions that can be found in \cite{Kobayashi:2006sb}. Notice that the 
$\sigma^3$ appearing in \eqq{Zdual} and \eqq{atdual} corresponds to the SU(2)$_\fla$ symmetry, not SU(2)$_\Rsym$.  An analogous expression holds with 
$\sigma^3$ replaced by the identity matrix. In view of this, the asymptotic behavior at large $r$
\be
\label{asym}
\phi(r) \simeq \Mq -  \frac{\Mq \Lambda^2}{r^2} + \cdots
\ee
has two immediate consequences. First,  
the constant $\Mq$ corresponds to the quark mass \cite{Karch:2002sh}.
To be precise, the quark mass is a complex number and we see that the masses of the $u$ and $d$ quarks are equal in magnitude but have opposite phases. Second, the corresponding quark condensates take the form 
\begin{subequations}
\begin{align}
\langle \overline \Psi \sigma^3 \Psi \rangle &= \langle \overline u u -  \overline d d \rangle \propto -  \Mq \Lambda^2 \,, \\[2mm]
\langle \overline \Psi \Psi \rangle &= \langle \overline u u +  \overline d d \rangle = 0  
\,,
\end{align}
\end{subequations}
or, equivalently:
\bea
\langle \overline u u \rangle = - \langle \overline d d \rangle \propto - \Mq \Lambda^2\,. 
\eea
The operator dual to the time component of the gauge field has $\Delta=3$ and reads schematically \cite{Kobayashi:2006sb}
\be
\label{atdual}
a_t  =\phi\,  \sigma^3 \,\,\,\,\,\, \leftrightarrow \,\,\,\,\,\, 
 J_t \, \sim \, \Psi^\dagger \sigma^3 \Psi \, \sim \, n_\mt{u} - n_\mt{d}  \,.
\ee
We have again omitted squark contributions that can be found in \cite{Kobayashi:2006sb}, and an analogous formula holds with $\sigma^3$ replaced by the identity matrix. Therefore the asymptotic behavior at large $r$
\be
\label{asymat}
a_t(r) \simeq \Mq -  \frac{\Mq \Lambda^2}{r^2} + \cdots
\ee
implies that the isospin chemical potential and the isospin, the $u$ and the $d$ charge densities are given by \cite{Apreda:2005yz,Kobayashi:2006sb}
\begin{subequations}
\label{through1}
\begin{align}
\label{muMq}
& \mui= \Mq \,, \\[2mm]
& \Ni = \Nu = -\Nd \propto - \Mq \Lambda^2 \,.
\end{align}
\end{subequations}
Finally, the operators dual to the magnetic components of the gauge field along the S$^3$ have $\Delta=2$ and are given by 
\cite{Kruczenski:2003be}
\be
\label{adual}
A_a = a(r) \, \delta_{a \, \hat a} \sigma^{\hat a} 
\,\,\,\,\,\, \leftrightarrow \,\,\,\,\,\, 
 {\cal O}_a = \psi^\dagger \sigma^{a} \psi \,,
\ee
where $\psi$ is the scalar superpartner of $\Psi$ and contains the squark components. Note that, unlike in \eqq{Zdual} and \eqq{atdual}, in \eqq{adual} the Pauli matrix on the right-hand side acts on the SU(2)$_\Rsym$ quantum numbers, not on the SU(2)$_\fla$ ones. It follows that the generic asymptotic behavior of the $a(r)$ field takes the form  
\be\label{eq.masslesslimit}
a \simeq  - \frac{v}{r^2} + \frac{s}{r^2} \log[r] + \cdots \,,
\ee
where $s$ and $v$ are the non-normalisable and the normalisable modes,  dual to the source and the vacuum expectation of ${\cal O}_a$, respectively.  In our case we see from \eqq{eq.solution} that the source vanishes and the VEV is given by the size of the instanton:
\be
v \sim \Lambda^2 \,.
\ee
We will show below that this is consistent with the fact that the non-trivial  $a(r)$ in our solutions breaks the color symmetry of the gauge theory spontaneously but not explicitly. If $\Mq=\mui=0$ this is familiar 
from the fact that in this case the ${\cal N}=2$ theory possesses a Higgs branch of vacua and $\Lambda$ is an exact modulus. In contrast, we will now see that in our case $\Lambda$ is fixed by other charges. 

We begin by noting that the presence of crossed electric and magnetic fields in the solution generates an angular momentum density. Inspection of the Poynting vector produced by these fields shows that the angular momentum density is aligned with the Cartan U(1)$_\lef$ subgroup of the SU(2)$_\lef$  preserved by our ansatz \cite{Eyras:2000dg}. In particular, this means that the angular momentum skew-symmetric two-form is self-dual with two identical skew-eigenvalues 
\be
\label{JJ}
J \equiv J_1=J_2 \,.
\ee
In our description the angular momentum arises as the conserved charge associated to the isometries of the S$^3 \subset \mathbb{R}^4$ and $n_1$ and $n_2$ are the corresponding angular momenta in two orthogonal planes of $\mathbb{R}^4$. From the viewpoint of the dual gauge theory, this corresponds to equal R-charge densities $n_1$ and $n_2$ along two of the three U(1) factors in the Cartan subalgebra of the SO(6) R-symmetry of  $\mathcal{N}=4$ SYM \cite{Gubser:1998jb,Chamblin:1999tk,Cvetic:1999ne}
\be
\label{through2}
\Nr \equiv n_1=n_2 \,,
\ee
with 
\be
\Nr \propto J \,.
\ee
In order to compute these charges,  let $\xi$ be the Killing vector associated to the U(1)$_\lef$ encoded in the left-invariant form $\omega_3$. Then the current $J_N=\xi^MT_{MN}$ is covariantly conserved, with $T_{MN}$ the energy-momentum tensor generated by the fields on the flavor branes. Integrating over the spatial volume gives the conserved angular momentum 
\begin{equation}
J=\int_{\Sigma}n^{N}J_N\,\d\Sigma \,,
\end{equation}
with $n^N=H^{1/4}\delta^N_t$ the unit normal to the constant-time slices $\Sigma$, whose volume element we denote $\d\Sigma$. A calculation then shows that the only non-vanishing component is 
\begin{equation}
\label{ang}
J=\int_{\Sigma}n^t\,T_{t3}\,\xi^3\,\d\Sigma  \propto
\int H^{1/4}\,\tr F_t{}^MF_{M3}\,\d\Sigma\propto\int_0^\infty r^3\left[a'\phi'+\frac{8}{r^2}a^2\phi\left(1+a\right)\right]\d r \propto  \Mq\Lambda^2\,.
\end{equation}
The third term in this expression makes it clear that the angular momentum is an integral of crossed electric and magnetic fields, as expected from the Poynting vector. We have omitted factors of the D7-brane tension for simplicity, as well as an overall (infinite) three-volume factor along the gauge theory directions. In other words, the result \eqq{ang} yields the angular momentum density per unit three-volume along the spatial 
$\vec{x}$-directions in  \eqq{eq.AdS5xS5metric}. The crucial point is that in \eqq{ang} all the factors of $H$, the harmonic function \eqq{eq.harmonicfunction},  cancel out exactly, so the result essentially reduces to its flat space counterpart. This property is a typical manifestation of the underlying supersymmetry preserved by the ground state  of the system. 

The analysis above shows that, up to normalisation, all charge densities in the system are comparable since they all scale as
\be
\label{still}
\Ni \sim \Nr \sim J \sim Q \sim \Mq \Lambda^2 \sim \mui \Lambda^2 \,.
\ee
Moreover, we see that the size of the instanton is not a free parameter but is actually given by 
\be
\label{lambda}
\Lambda^2 \propto \frac{n_\Rsym}{\mui} \,.
\ee
This can be understood in simple mechanical terms \cite{Eyras:2000dg}: The non-vanishing angular momentum $J\sim \Nr$ prevents the instanton from collapsing to zero size despite the fact that the non-Abelian gauge symmetry on the D7-branes is broken to an abelian subgroup as described in \eqq{implement}. Alternatively, the stability of the solution can be understood  \cite{Lambert:1999ua} from the dyonic nature of the instanton, namely from the fact that it is precisely the breaking of the gauge symmetry to an Abelian subgroup that allows for the definition of the electric charge of the instanton via the projection \eqq{QQQ}.

To close this circle of ideas it is interesting to compute the total energy of the D7-branes, since this will bring about a covariant expression for the electric charge \eqq{QQQ}. In this case we contract the stress tensor with the timelike killing vector in the geometry $\xi=\partial_t$ to obtain
\begin{equation} 
\mathcal{E}=\int_{\Sigma}n^t\,T_{tt}\,\xi^t\d\Sigma\propto\int\d^4y\,\tr\left(\frac14\delta^{ij}\delta^{kl}F_{ik}F_{jl}+\delta^{ij}E_iE_j\right)\,,
\end{equation}
where again we have omitted   an overall three-volume factor and, crucially, all factors of $H$ have cancelled out. Using self-duality of the solution, the first term is proportional to the instanton charge \eqq{eq.instantonnumber}. The second term is the energy stored in the electric field. Using \eqq{deri} and integrating by parts this evaluates to
\begin{equation}
\label{total}
\int\d^4y\,\tr\delta^{ij}E_iE_j \propto
\lim_{r\to\infty} \int_{S^3} r^3 \, \tr\! \left( Z E_r \right) 
\propto  \Mq^2 \Lambda^2 = \Mq Q\,.
\end{equation}
Thus the energy can be written exclusively in terms of conserved charges as 
\be
\mathcal{E} \sim k + \Mq Q \,,
\ee
showing that it saturates a BPS bound \cite{Lambert:1999ua}.

\subsection{Infrared limit} 
\label{IRlimit}
 In this limit $r\to 0$ the fields approach the values 
\be\label{eq.IRsolution}
a_t=\phi \to 0 \sac a \to -1 \ .
\ee
Under these circumstances the metric on the D7-branes approaches AdS$_5\times$S$^5$ with the same radius as in the UV and the non-Abelian field strength tends to zero. This configuration describes a fixed point and is actually a solution by itself, i.e.~it can be detached from the RG flow that ends on it. The fact that the IR radius is the same as in the UV  would seem to suggest that the physics near the IR fixed point is the same near its UV counterpart. However, this is not the case because the IR value $a=-1$ is physically inequivalent to the UV value $a=0$, since the two configurations differ by a large gauge transformation \cite{Erdmenger:2005bj}. This can be seen by considering fluctuations of the different fields around the $a=-1$ solution. At quadratic order, the masses of these fluctuations reveal that the time-component of the gauge field, $a_t$, and the scalar, $Z$, are dual to $\Delta=5$ operators \cite{Arean:2007nh}, whereas the field $a$ is dual to a $\Delta=6$ operator. These dimensions imply that the generic behavior of fluctuations around the IR fixed point is
\be
\label{deltadelta}
\delta a_t= \delta \phi(r) \simeq v_5\, r^{-4} + s_5\, r^2 \ , \qquad \delta a(r) \simeq v_6 \, r^{-6} + s_6\, r^2 \,,
\ee
where $s_\Delta$ and $v_\Delta$ stand for the source and VEV of the operator with dimension $\Delta$. The fall-off in the IR of the fields in our solution takes the form 
\be
a_t(r) = \phi(r) \simeq 0 + \frac{\Mq}{\Lambda^2} \, r^2 \sac
a(r) = -1 + \frac{1}{\Lambda^2} \, r^2 \,.
\ee
Comparing with \eqq{deltadelta} we see that the VEVs vanish along our RG flow, which therefore describes a deformation of the IR fixed point by irrelevant operators.

\subsection{Spontaneous symmetry breaking \label{sec.symmetrypatterns}}
In \Sec{modelsec} we discussed the explicit breaking of symmetries, first by coupling the original  ${\cal N}=4$ SYM theory to dynamical quarks, and then by the non-zero mass of these quarks in the resulting  ${\cal N}=2$ theory. In this section we will discuss the spontaneous breaking of some of the remaining symmetries by the presence of the instanton field. 

\begin{figure}[t]
\begin{center}
\includegraphics[width=.65\textwidth]{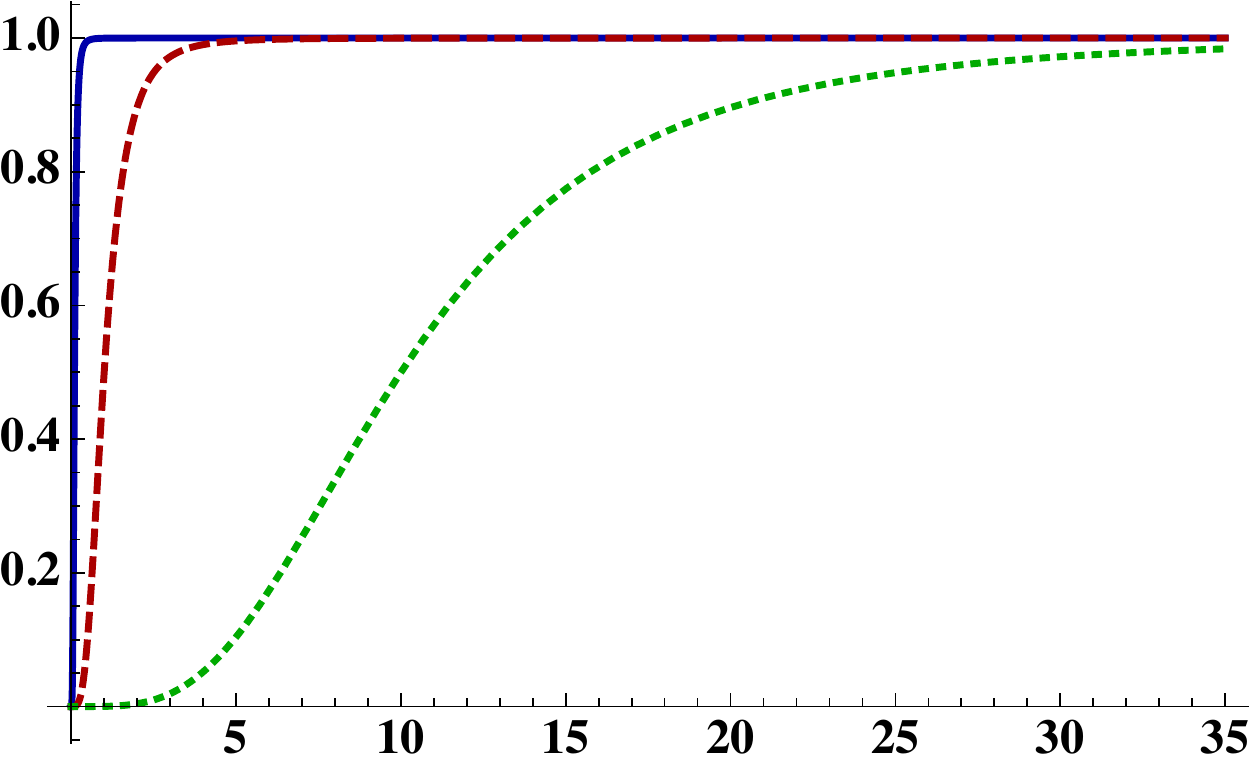}
\put(-305,190){\Large $\nc(r)-\nc $}
\put(5,15){\Large $r/\Mq$}
\end{center}
\vspace{-1mm}
\caption{\small Effective number of D3-branes on the D7-branes within a sphere of radius $r$ --- see \eqq{effective} ---  for 
$\Lambda/\Mq=\{ 1/10,1,10\}$ (solid blue, dashed red and dotted green curves, respectively).\label{fig.numberofbranes}}
\end{figure}

Let us first consider the color  symmetry of the gauge theory. This is broken spontaneously as 
\be
\textrm{SU}(\nc+1) \to \textrm{SU}(\nc) \times \textrm{U}(1) 
\ee
due to the dissolution of a unit of D3-brane charge inside the D7-branes. Intuitively, one can understand this from the fact that the dissolved D3-brane has been separated from the rest, hence breaking or Higgsing the color gauge group. This breaking is spontaneous because the instanton field is normalisable. In other words, the dissolution of the D3-branes on the D7-branes is not forced by an external force for the instanton field but is induced by the other charges in the system. This is apparent from \eqq{lambda}, which shows that the instanton size is not an independent quantity but is fixed by the other charges. The fact that the instanton field carries D3-charge follows simply from the coupling 
\be
\int C_4 \wedge \tr \left( F \wedge F \right)
\ee 
on the worldvolume of the D7-branes, which shows that the instanton density couples to $C_4$ in the same way that D3-branes do. As a consequence of the extended nature of this density, the effective number of D3-branes in our solution contained within a sphere of radius $r$ on the D7-branes is given by --- c.f.~Eq.~\eqq{eq.instantonnumber}:  
\be
\label{effective}
\nc(r)=\nc -  \Big[ 3 a^2 + 2 a^3 \Big]^{r}_{0} =
\nc + 1 - 3\, a(r)^2 - 2\, a(r)^3 \,.
\ee
The quantity $\nc(r)-\nc$ is plotted in Fig.~\ref{fig.numberofbranes}.

The instanton field also breaks spontaneously part of the global symmetries as 
\be
\label{implied}
{\mbox{SU(2)}_\mt{R}\times\mbox{U(1)}_\iso
\to\mbox{U(1)}_\mt{D}} 
\,,
\ee
where the group on the right-hand side is the diagonal U(1) in the Cartan subalgebra of the left-hand side. This breaking can be seen geometrically in two steps. First, in the presence of the  isospin electric field, the magnetic components that are turned on by the instanton generate a self-dual angular momentum in $\mathbb{R}^4$ given by \eqq{ang}. Generically, a self-dual two-form in $\mathbb{R}^4$ is invariant  only under the $\mbox{SU(2)}_\lef \times \mbox{U(1)}_\Rsym$ subgroup of SO(4). In other words, it breaks 
$
\mbox{SU(2)}_\Rsym \to  \mbox{U(1)}_\Rsym \,. 
$
In our case this  manifests itself in the fact that the Poynting vector distinguishes between the $\omega^3$ and the $\omega^{1,2}$ forms that enter the ansatz  \eqq{eq.ansatz}, thus breaking $\mbox{SU(2)}_\mt{R}$ to the $\mbox{U(1)}_\mt{R}$ subgroup that rotates $\omega^{1,2}$ into one another. Second, the further breaking 
\be
{\mbox{U(1)}_\mt{R}\times\mbox{U(1)}_\iso
\to\mbox{U(1)}_\mt{D}} 
\ee
implied by \eqq{implied} is due to the fact that only simultaneous rotations of $\omega^a$ and $\sigma^{\hat a}$ leave the last term in \eqq{eq.ansatz} invariant. This is similar to the breaking 
$
\mbox{SU(2)}_\mt{R} \times \mbox{SU(2)}_\fsub \to 
\mbox{SU(2)}_\mt{D}
$
that takes  place on the Higgs branch  in the absence of the isospin electric field (see e.g.~\cite{Erdmenger:2005bj}). Intuitively, it is simply due to the fact that an instanton configuration in $\mathbb{R}^4$ is only invariant under simultaneous $\mbox{SU(2)}_\mt{R}$ rotations in space and 
$\mbox{SU(2)}_\fsub$ large gauge transformations that act on the orientation modes of the instanton. This is a well known property that is crucial for the correct counting of the number of instanton zero modes.

\section{Spectrum\label{sec.fluctuations}}

In this section we will investigate the spectrum of mesonic excitations around the ground state that we have described above. To do so we will consider linear fluctuations around the solution. For simplicity,  we will assume that their dynamics is governed by the SYMH action \eqref{eq.ourmodel}  instead of the action \eqref{wvaction}. Although this may affect some quantitative details of the spectrum, we expect that the qualitative properties will be the same, in particular those related to Goldstone modes.

In principle, fluctuations can depend on all  the coordinates of the D7-branes, $\{t,\vec x, r, \theta, \phi, \psi \}$, with $\{\theta, \phi, \psi \}$ the angles of the $S^3$. However, instead of working with generic dependence on these angles, we will restrict the analysis to the angular dependence encoded in the left-invariant one-forms $\omega^a$ that we used in the ansatz for the background \eqref{eq.ansatz}. This 
amounts to a restriction to the $\mbox{SU(2)}_\lef$-invariant sector of the spectrum. In particular, this means that the fluctuations that we will consider will keep the centre of the instanton at the origin of AdS$_5$, only allowing for fluctuations of its size $\Lambda$. In \Sec{sec.moreungapped} we will comment on how to relax this restriction.

Including the fluctuations, we consider the following form for the scalar fields 
\be\label{eq.Zfluc}
Z_1 = \phi(r) \, \sigma^3 + \zeta_{\hat a}(t,\vec x,r) \, \sigma^{\hat a} \ , \qquad Z_2 = \beta_{\hat a}(t,\vec x,r) \, \sigma^{\hat a} \ ,
\ee
and for the gauge field 
\be\bal\label{eq.Afluc}
A & = a_t(r) \d t \otimes \sigma^3 + a(r) \delta_{a {\hat a}}  w^a \otimes \sigma^{\hat a} \\[2mm]
& \quad + \Big[ \alpha_{\mu {\hat a}}(t,\vec x,r)\, \d x^\mu   +  \alpha_{r {\hat a}}(t,\vec x,r)\, \d r  +  \alpha_{a {\hat a}}(t,\vec x,r)\, w^a \Big] \otimes \sigma^{\hat a} \ .
\eal\ee
Note that we have focused on fluctuations inside SU(2)$_\fla$ and have omitted those along the U(1)$_\bar$ subgroup of U(2)$_\fla$, which would be proportional to $\sigma^0$. From now on we fix the holographic gauge 
\be
\label{fix}
\alpha_{r {\hat a}}=0 \,.
\ee
We also work in momentum space and use the SO(3) rotations along the gauge theory spatial directions to align the momentum $k$ with one of these coordinates, which we denote simply as $x$. The fluctuations split into seven different channels according to their behavior with respect to large gauge transformations and  to the  global symmetry group SO(2)$\times$U(1)$_\mt{D}$, where SO(2)  is the little group with respect to the momentum. The charges of the different fluctuations (divided by 2 to simplify the notation) under the U(1)$_\iso$ and U(1)$_\Rsym$ groups are given in Table \ref{tab.charges}, where we have defined
\be
\alpha_\pm = \alpha_{11} + \alpha_{22} \pm i ( \alpha_{12} - \alpha_{21} ) \ , \qquad  \widehat \alpha_\pm = \alpha_{11} - \alpha_{22} \pm i ( \alpha_{12} + \alpha_{21} ) \ .
\ee
\begin{table}[t]
\begin{tabular}{c|ccccccccccc}
 & $\alpha_{\mu1} \pm i \alpha_{\mu2}$ & $\alpha_\pm$ &  $\widehat \alpha_\pm$  & $\alpha_{31}\pm i \alpha_{32}$ & $\alpha_{13} \pm i \alpha_{23}$ & $\beta_1 \pm i \beta_2$ & $\zeta_1 \pm i \zeta_2$ & $\alpha_{\mu 3}$  \\
\hline
U(1)$_\iso$ & $\pm$ & $\pm$ & $\pm$ & $\pm$ & $0$ & $\pm$ & $\pm$ & $0$  \\
U(1)$_\Rsym$ & $0$ & $\mp$ & $\pm$ & $0$ & $\pm$ & $0$ & $0$ & $0$ \\
\hline
gauge & $\lambda_{1,2}$ & $\lambda_{3}$ &  & $\lambda_{1,2}$ & $\lambda_{1,2}$ &  & $\lambda_{1,2}$ & $\lambda_{3}$ & 
\end{tabular}
\caption{\small Charges (divided by two) of the fluctuations under the relevant Abelian global symmetries.\label{tab.charges}}
\end{table}

Residual gauge transformations  that preserve the condition \eqq{fix} take the form
\be
\label{gaugetrafo}
A \to A + \d \lambda + [A, \lambda] \ , \quad  
Z^\alpha \to Z^\alpha + [ Z^\alpha,  \lambda] 
\ee
with
\be
 \lambda =  \lambda_{\hat a}(t,\vec x)\, \sigma^{\hat a} \ .
\ee
Note that these are large gauge transformations because the gauge parameter $\lambda$ does not approach the identity at the boundary, since it is $r$-independent. Explicitly, the action of these transformations on the background fields reads
\be\label{eq.gaugetransformation}
\bal
\d \lambda_{\hat a} & =   \left( - i \, \omega\, \d t + i \, k \, \d x \right) e^{-i \omega t + i k x}  \lambda_{\hat a}  \ , \\[2mm]
[A_\mt{b.g.},\lambda]_{\hat a} & = 2\,  \left( - a_t(r) \delta_{a \hat a}\,\delta_{b\hat b}\, \epsilon^{ab3}  \,\d t + a(r) \delta_{c \hat c} \, \delta_{\hat a \hat d}\, \epsilon^{{\hat d}{ \hat c}{\hat b}}   \,  \omega^{c}  \right) e^{-i \omega t + i k x} \lambda_{\hat b} \ , \\[2mm]
[Z^1_\mt{b.g.},\lambda]_{\hat a} & = - 2\, \phi(r) \delta_{a \hat a} \, \delta_{b\hat b} \, \epsilon^{ab3} \, e^{-i \omega t + i k x} \lambda_{\hat b}  \ .
\eal
\ee
These expressions show that some combinations of fluctuations can be generated by acting on the background solution with a large gauge transformation. This is summarised in the last row of Table \ref{tab.charges}. For example, the entries with a $\lambda_3$ mean that if $\lambda_3 \neq 0$ then acting on the background \eqq{eq.ansatz} with the gauge transformation \eqq{gaugetrafo} generates a new background with non-zero terms of the form parameterised by $\alpha_\pm$ and $\alpha_{\mu3}$ in \eqq{eq.Afluc}.

The SO(2) scalars $\alpha_{3 3}$, $\beta_3$ and $\zeta_3$ are not included in the table because they are neutral under both U(1)$_\iso$ and U(1)$_\Rsym$ and they cannot be generated from the background by a residual gauge transformation.

A general and detailed discussion of all the different channels, including the asymptotic behavior of the fields and our numerical procedure to solve their equations of motion, is given in  Appendix \ref{app.fluctuations}. In contrast, in the next sections we will focus on some particularly interesting sectors.

\subsection{Goldstone modes\label{sec.goldstones}}

The spontaneous breaking of the global symmetries \eqq{implied} suggests that the spectrum should contain ungapped, Goldstone  modes. The channels where we expect them to arise can be identified by considering the action of the broken generators on the ground state that breaks them. For infinitesimal transformations this action should generate a family of inequivalent ground states with the same energy that differ from the original one by a small perturbation. At zero energy and zero momentum, this perturbation is normalisable and corresponds precisely to the ungapped mode. 

In our case, the group SU(2)$_\Rsym \times$U(1)$_\iso$ acts as 
\be\label{eq.explicitsymmetry}
\begin{pmatrix} \omega^1 \\ \omega^2 \\ \omega^3 \end{pmatrix} \to \mathbf{R}_1(\vartheta) \cdot \mathbf{R}_2(\varphi) \cdot \mathbf{R}_3(\psi) \cdot \begin{pmatrix} \omega^1 \\ \omega^2 \\ \omega^3 \end{pmatrix} 
 \ , \qquad
\begin{pmatrix} \sigma^1 \\ \sigma^2 \\ \sigma^3 \end{pmatrix} \to   \mathbf{R}_3(\delta) \cdot \begin{pmatrix} \sigma^1 \\ \sigma^2 \\ \sigma^3 \end{pmatrix} \,.
\ee
In this expression the angles $\vartheta$, $\varphi$, $\psi$ and $\delta$ can depend on the spacetime coordinates $t,x,r$ and $R_i$ is a $3\times 3$ rotation matrix around the $i$-th axis. For example, $R_3(\psi)$ rotates $\omega^1$ and $\omega^2$ into one another. The solution \eqref{eq.ansatz} is invariant under the subset of these transformations that obeys 
\be
\varphi = \vartheta =0 \sac  \delta - \psi = 0\,, 
\ee
which corresponds precisely to the preserved U(1)$_\mt{D}$ symmetry. Consider thus a transformation of the form \eqq{eq.explicitsymmetry} that does not obey this condition. In the zero-momentum limit, the action of this transformation on the solution \eqref{eq.ansatz} generates a new,  inequivalent ground state. A calculation shows that the new solution is of the form \eqq{eq.Zfluc}-\eqq{eq.Afluc} where the only non-zero extra components are the following skew-symmetric components in the $\alpha_{a \hat a}$ matrix:
\begin{subequations}
\label{both}
\begin{align}
\label{both1}
& \alpha_{12} - \alpha_{21} = 2 (\delta-\psi)\, a(r)
 \,, \\[2mm]
 \label{both2}
& \alpha_{13} - \alpha_{31} = - 2 \varphi \, a(r)\sac
\alpha_{23} - \alpha_{32} = - 2 \vartheta \, a(r)\,.
\end{align}
\end{subequations}
Comparing \eqq{both1} with  \eqref{eq.gaugelambda3} and \eqq{both2} with \eqref{eq.gaugelambda12} we see that this transformation generates an $\alpha_{a \hat a}$ matrix equivalent to that generated by large gauge transformations provided we identify 
\be
\delta-\psi = 2 \lambda_3 \ , \qquad \varphi = 2 \lambda_2 \ , \qquad \vartheta = -2 \lambda_1 \,.
\ee
This is no surprise because the last term in \eqref{eq.ansatz} (but not the others) is actually invariant under the full SU(2)$_\mt{D}$ subgroup of 
SU(2)$_\Rsym\times$SU(2)$_\fla$, which means that an SU(2)$_\Rsym$ rotation is equivalent to an SU(2)$_\fla$ large gauge transformation.

\begin{figure}[t!!!]
\begin{center}
\includegraphics[width=.6\textwidth]{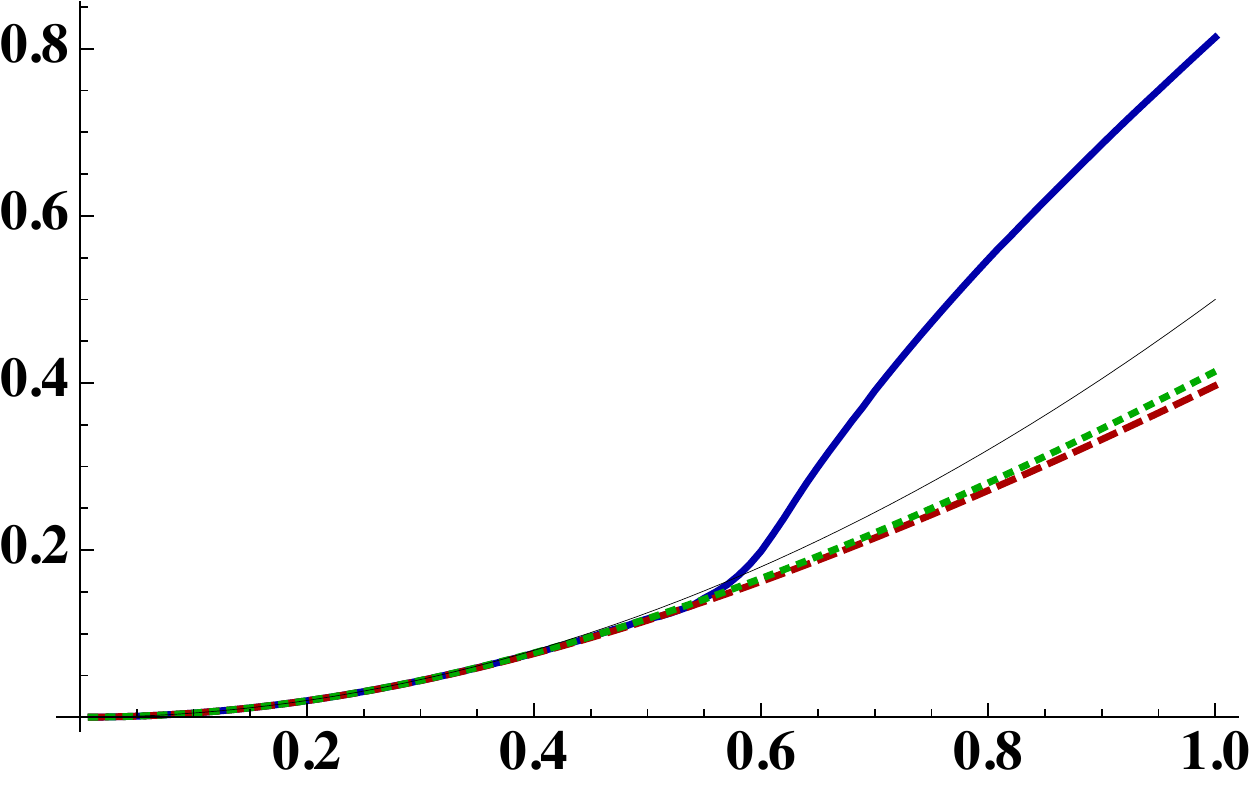} 
\put(-265,178){\Large \mbox{ $\omega/\mui$}}
\put(2,13){\Large \mbox{ $k/\mui$}}
\\[15mm]
\includegraphics[width=.6\textwidth]{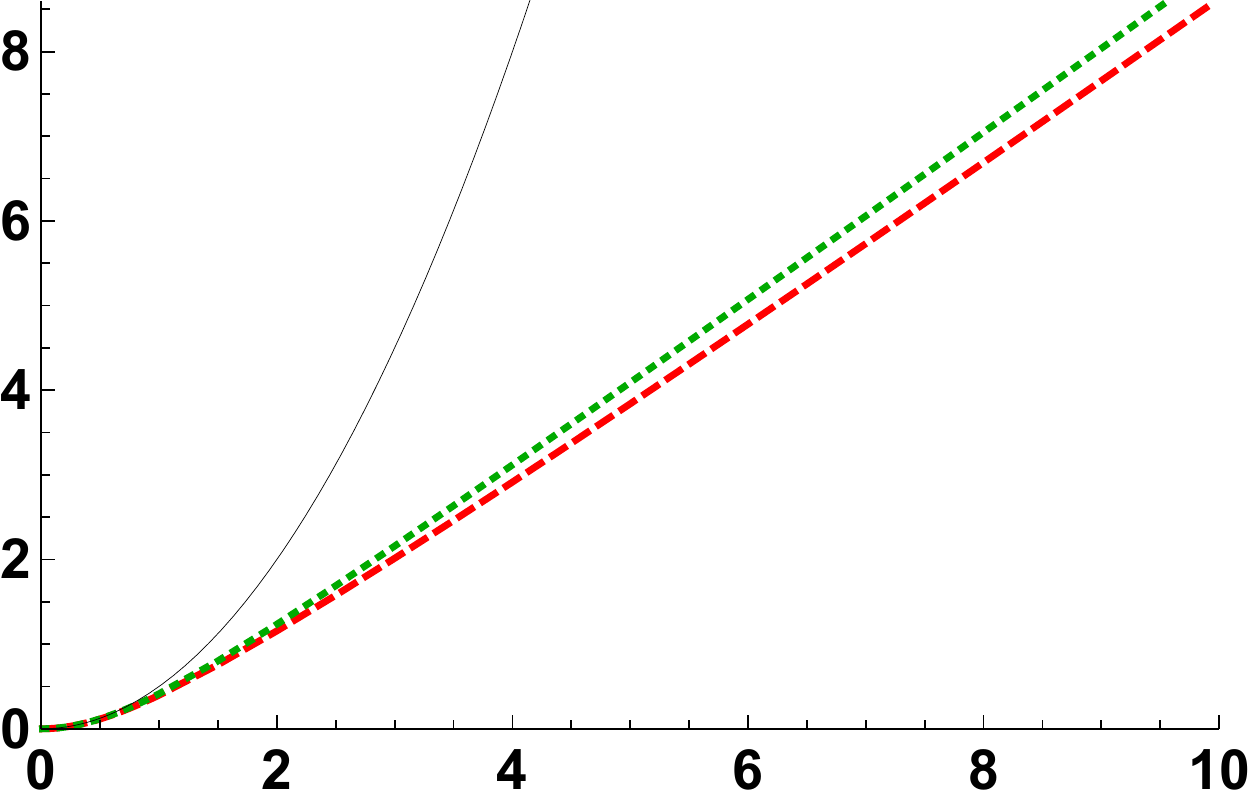} 
\put(-265,178){\Large \mbox{ $\omega/\mui$}}
\put(2,13){\Large \mbox{ $k/\mui$}}
\end{center}
\vspace{-1mm}
\caption{\small Dispersion relation of the Goldstone bosons for 
$ \Lambda/\Mq=1/10$ (solid, blue curve), $\Lambda/\Mq=1$ (dashed, red curve), and $\Lambda/\Mq=10$ (dotted, green curve) at small (top) and large (bottom) momenta. The thin black line corresponds to the low-momentum behavior \eqref{eq.Goldstones} and we have set $L=1$ in this integration.
\label{fig.goldstones}
}
\end{figure}

The modes of the form \eqq{both1} and \eqq{both2} appear in the channels number 4 and 6 in Appendix \ref{app.fluctuations}, respectively. This suggests that the ungapped modes associated to the breaking \eqq{implement} should appear in these channels. Our numerical analysis, explained in detail in Appendix \ref{app.fluctuations}, confirms this. In each channel we find a gapless, non-dissipative quasi-normal mode. In other words, a mode with strictly real  frequency that vanishes at zero-momentum. The dispersion relation at low momentum is given by
\be\label{eq.Goldstones}
\omega = \pm \frac{1}{2 \mu_\iso} k^2 \,,
\ee
showing that these modes are also non-relativistic. It is  interesting that the low-momentum behavior is independent  of the 't Hooft coupling and of the dimensionless ratio $\Lambda/\Mq$. At larger momenta the dispersion relation is modified and approaches the form
\be
\omega = \omega_0 \left( \lambda, \frac{\Lambda}{\Mq} \right) + k \,, 
\ee
as illustrated in  Fig.~\ref{fig.goldstones}. We emphasise that these dispersion relations are the same for the Goldstone modes in both channels. 
Our numerical analysis has not allowed us to establish whether the Goldstone mode in channel 6 is a single mode or actually corresponds to two exactly degenerate modes. However, \eqn{both} shows that the transformations  parametrised by $\varphi$ and $\vartheta$ generate different solutions. This strongly suggests that  the mode we find in channel 6 is indeed doubly degenerate, in which case the number of Goldstone modes with a non-relativistic dispersion relation, also known as type II Goldstone modes, would be  $n_\mt{II}=3$.

Let us now place our results on the general context of theorems about the existence of ungapped modes, following a similar discussion in \cite{Amado:2013xya}. We first recall that Goldstone's theorem only implies that  the number of ungapped modes is equal to the number of broken generators, $N_\mt{BG}$, in the presence of Poincar\'e symmetry. In this case the ungapped modes are called type I modes because they must have a relativistic dispersion relation at low momentum, 
$\omega \sim k$, and we have that 
\be
n_\mt{I}=N_\mt{BG} \,.
\ee
 If Lorentz invariance is broken, for example because of a non-zero charge density, then some modes may have a type II dispersion relation, meaning that $\omega \sim k^2$ at low momentum. In this case the number of modes and the number of broken generators obeys the Chadha and Nielsen inequality \cite{Nielsen:1975hm}
\be\label{eq.countinggoldstones}
n_\mt{I} + 2 \, n_\mt{II} \geq N_\mt{BG} \,.
\ee
The number of type I and type II Goldstone bosons can be further constrained. If the broken symmetry generators obey 
\be
\langle [Q_a,Q_b] \rangle = B_{ab} \,,
\ee
 then the number of Goldstone bosons should satisfy  \cite{Watanabe:2011ec,Watanabe:2012hr,Watanabe:2013iia} 
\be
\label{eq.countinggoldstonesBIS}
n_\mt{I} +  n_\mt{II} = N_\mt{BG} - \frac{1}{2} \mbox{rank} (B )\,.
\ee
In our case the number of broken generators needs some discussion. If we were to view all the generators in  \eqq{implied} on an equal footing then the broken generators would be $\tau^1, \tau^2$ and  
$\tau^3-\sigma^3$, with $\tau^n$ the generators of 
$\mbox{SU(2)}_\mt{R}$. Therefore we would have $N_\mt{BG}=3$ and the antisymmetric matrix $B$ would have rank 2 with entries 
\be
B_{12} \sim J \sac B_{13}=B_{23}=0 \,.
\ee
It would then follow that \eqq{eq.countinggoldstones} would be satisfied whereas \eqq{eq.countinggoldstonesBIS} would not. However, in our case
the symmetries SU(2)$_\Rsym$ do not have an associated conserved current in the dual gauge theory, which is an assumption in the theorems quoted above. The reason is that, on the gravity side, the gauge fields associated to the SU(2)$_\Rsym$ symmetries are off-diagonal components of the metric along time and one of the S$^3$ angular directions. In our probe approximation these fields are non-dynamical since we ignore the backreaction of the branes on the spacetime metric. In the dual gauge theory this statement translates into the fact that there is no conserved current that implements the SU(2)$_\Rsym$ symmetry, which is therefore better regarded as an ``outer automorphism'' of the operator algebra of the gauge theory. From this viewpoint there are no  broken generators, the matrix $B$ trivially vanishes and  \eqq{eq.countinggoldstones} is satisfied but  \eqq{eq.countinggoldstonesBIS}  is not.

\subsection{Pseudo-Goldtstone modes at $\Mq \ll \Lambda$}

\begin{figure}[t]
\begin{center}
\includegraphics[width=.45\textwidth]{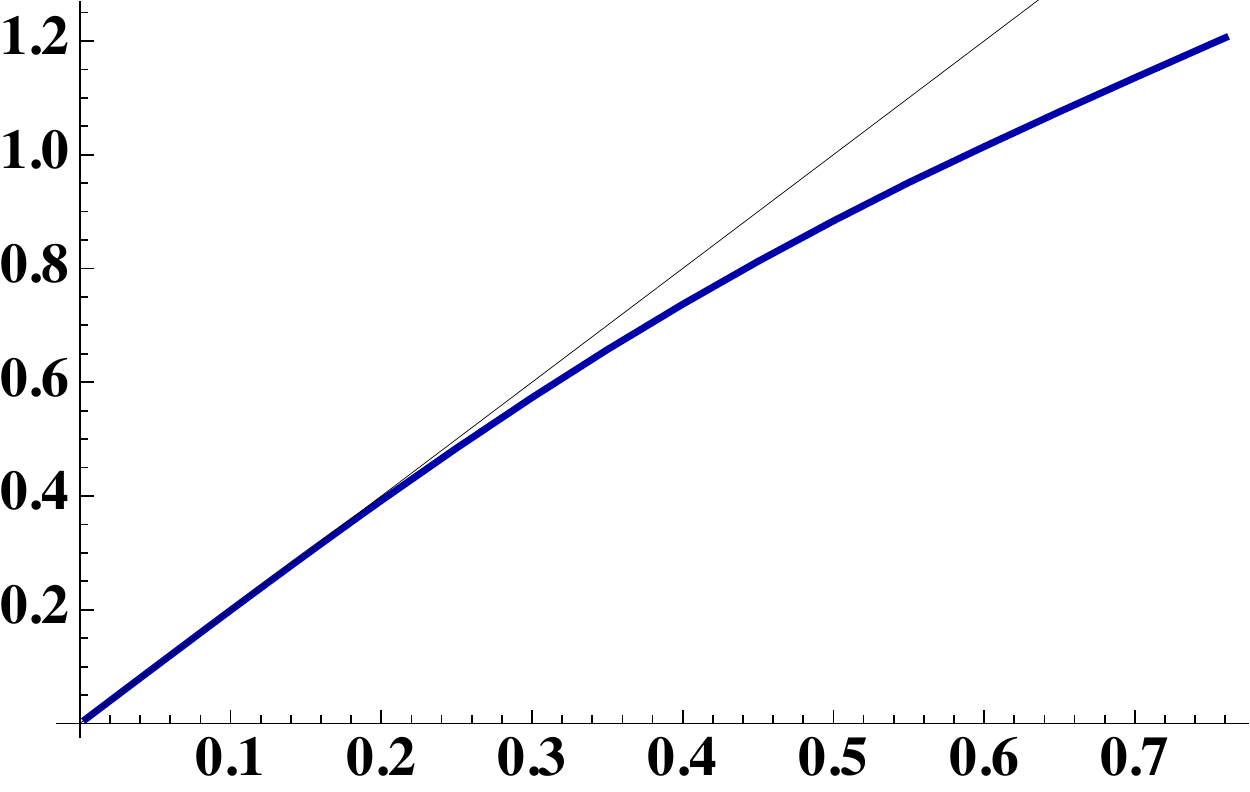} 
\put(-210,145){\large \mbox{ $\omega_\mt{pseudo}/\Lambda$}}
\put(2,5){\mbox{ $M_q/\Lambda$}}
\end{center}
\vspace{-1mm}
\caption{\small Mass of the pseudo-Goldstone mode. The thin, black line corresponds to the relation $\omega_\mt{pseudo} =2 M_q$.
\label{fig.pgbmass}
}
\end{figure}

The physics in our system is controlled by the ratio of the two dimensionful  scales and by the 't Hooft coupling 
\be
\overline \Lambda \equiv \frac{\Lambda}{\Mq} = \frac{\Lambda}{\mu_\iso}  \  , \qquad \lambda = \frac{L^4}{ 2\pi^2} \ .
\ee
In the limit $\Mq=\mui \to 0$ with fixed $\Lambda$ the size of the instanton becomes an exact modulus, so we expect the presence of an ungapped excitation in the system associated to the possibility of shifting 
\be
\Lambda \to \Lambda + \delta \Lambda 
\ee
with no associated energy cost. Indeed,  if we set 
$\Mq=\mui = 0$  then there is a simple solution to the fluctuation equations of motion of Sec.~\ref{app.gaugelambda3} given by
\be\label{eq.goldstonescaling}
\alpha_{33}=\frac{1}{2}\alpha_\pm = \delta \Lambda \, \partial_\Lambda a(r) \ , \quad \zeta_3 = \alpha_{t3} = \alpha_{x3}= 0 \ , \quad \omega^2 = k^2  \,,
\ee
where $\delta \Lambda$ is an arbitrary integration constant and 
$\partial_\Lambda a(r)$ is the derivative of the solution \eqq{ar} with respect to the parameter $\Lambda$, which automatically produces a normalisable mode.   Furthermore, the equations of motion force the dispersion relation
\be
\omega^2=k^2 \ ,
\ee
corresponding to  a relativistic Goldstone mode associated to the spontaneous breaking of scale invariance by a non-zero $\Lambda$. Turning on a non-zero quark mass and chemical potential such that $\Mq = \mui\ll \Lambda$ introduces a small  amount of explicit breaking of scale invariance, so in this case we expect the Goldstone mode above to become a pseudo-Golsdstone mode with a small mass proportional to $\Mq = \mui$. To verify this we perform a numerical integration of the equations of motion at zero spatial momentum  and scan the real frequency axis as a function of $\Mq/\Lambda$.
From the numerical results we conclude that there exists  a  mode whose frequency in the limit $\Mq \ll \Lambda$ is given by 
\be\label{eq.pseudogoldstone}
\omega_\mt{pseudo} \simeq 2 \Mq\,.
\ee
This behavior can be seen  in \fig{fig.pgbmass} for low values of $\Mq$.  
The figure also shows that the gap ceases to be linear in $\Mq$ as the quark mass is increased and  thus the scales of explicit and spontaneous symmetry breaking become comparable to one another. Within our numerical precision the frequency remains purely real for all values of $\Mq / \Lambda$, meaning that this mode is absolutely stable. 

It is interesting to note that, under certain circumstances, non-relativistic Goldstone modes may be accompanied by massive modes whose mass is proportional  to the scale of Lorentz-symmetry breaking \cite{Nicolis:2012vf,Kapustin:2012cr}. In our case the relationship would be 
\be\label{eq.almostgoldstone}
\omega=q\, \mu_\iso \,,
\ee
with $q$ the charge of the fluctuations under the isospin group. Since 
$\Mq=\mu_\iso$ and in our conventions $q=2$,  this value coincides with \eqref{eq.pseudogoldstone}. It is therefore tempting to interpret the pseudo-Goldstone mode \eqref{eq.pseudogoldstone} as the partner of  the Goldstone modes of the previous section. However, there seems to be no possible match in terms of the number of modes. Although it is difficult to establish numerically whether or not the  pseudo-Goldstone mode is degenerate, the  $\Mq=\mui = 0$ result \eqq{eq.goldstonescaling} suggests that there is only one of these modes, in contrast to the three Goldstone modes of the previous section.

\subsection{Massive quasiparticles at $\Lambda \ll \Mq$ 
\label{sec.massivespectrum}}

\begin{figure}[t]
\begin{center}
\includegraphics[width=.6\textwidth]{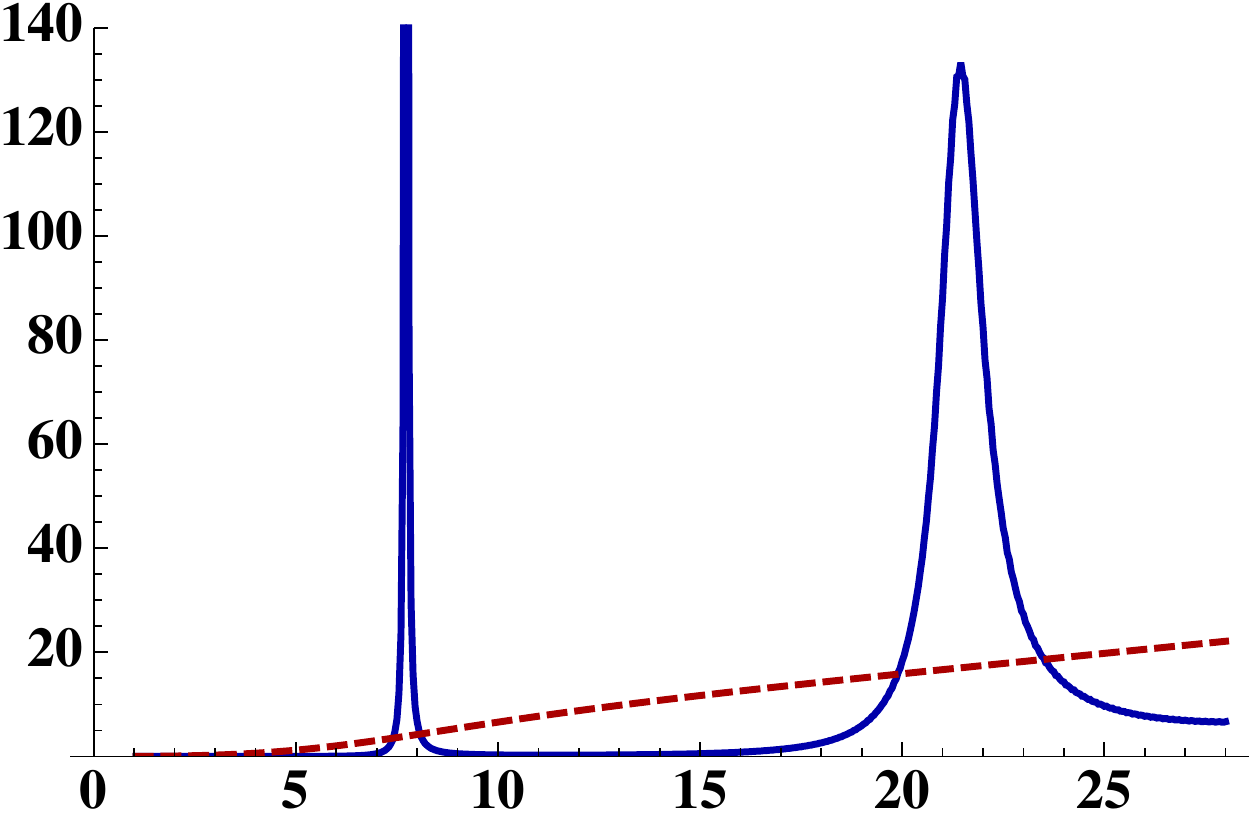} 
\put(-285,90){\mbox{\Large $\chi$}}
\put(-148,-18){\mbox{\large $L^4 \omega^2 / M_q^2$}}
\end{center}
\vspace{-1mm}
\caption{\small Spectral function (with arbitrary normalisation) of the gauge theory operator dual to the $\beta_3$ scalar  for  $\overline \Lambda=1$ (dashed, red curve) and $\overline \Lambda=1/10$ (solid, blue curve).
}
\label{fig.spectralfunction}
\end{figure}

As illustrated in  Fig.~\ref{fig.profile}, in the limit  $\overline \Lambda \ll 1$ the embedding of the branes approaches that of two parallel branes connected only at the origin by a very thin throat. Therefore one may expect that the spectrum should approach that of Ref.~\cite{Kruczenski:2003be}, at least for the 
U(1)$\iso$-neutral fields $\beta_3$, $\zeta_3$, $\alpha_{33}$, $\alpha_{\mu\, 3}$, and ${\alpha_{13}\pm i\, \alpha_{23}}$ (see Table  \ref{tab.charges}). The spectrum of \cite{Kruczenski:2003be} is
\be\label{eq.massspectrum}
\omega^2 - k^2 = \frac{4\Mq^2}{L^4}(n+1)(n+2) = 8 \pi^2 \frac{\Mq^2}{\lambda} (n+1)(n+2) \,,
\ee
where $n$ is the radial quantum number. Note that these frequencies are purely real since in the limit $\overline \Lambda=0$ there is no possible absorption of these modes by the horizon at leading order in the large-$\nc$, large-$\lambda$ expansion.  If instead $\overline \Lambda$ is very small but non-zero we expect that each of these modes will develop a small, negative imaginary part since now the branes reach the Poincar\'e horizon along a thin throat. In other words, for small but finite $\overline \Lambda$ the normal modes of \cite{Kruczenski:2003be} become quasi-normal modes  located in the complex lower  half-plane near the real axis. This manifests itself in high and narrow peaks in the  spectral function of the corresponding dual operator. This is illustrated in  \fig{fig.spectralfunction} for the case of the operator dual to  the $\beta_3$ scalar. As one increases 
$\overline \Lambda$ the peaks cease to exist. The numerical procedure used to produce \fig{fig.spectralfunction} is explained in  \Sec{app.beta3fluctuation}. 

Fluctuations charged under U(1)$_\iso$ have a different spectrum  because they couple to the isospin chemical potential. Comparing for example Eqn.~\eqref{eq.beta12eoms} to Eqn.~\eqref{eq.beta3eom} it is easy to see  that, in the limit $\overline \Lambda=0$, these fluctuations have normalisable modes at frequencies given by 
\be\label{eq.isospinchargedspectrum}
\left( \omega \pm 2 \mu_\iso \right)^2 - k ^2 = 4 \Mq^2 + 8\pi^2\, \frac{\Mq^2}{\lambda} (n+1)(n+2) \,,
\ee
i.e. the frequencies are  shifted due to the presence of the finite isospin chemical potential, with the factor of $2$ given by the isospin charge of the fluctuations. Note that in the limit $\overline \Lambda=0$ one can turn on an isospin chemical potential while keeping the isospin charge to zero (and this configuration will be dominant for $\mui<\Mq$). Under these circumstances the chemical potential does not affect the embedding profile of the branes but it breaks the mass degeneracy of isospin multiplets
\cite{Erdmenger:2007ap,Aharony:2007uu,Erdmenger:2007ja}. In the case of triplets one mode is unaffected, as in \eqq{eq.massspectrum}, one mode increases in mass, as in \eqq{eq.isospinchargedspectrum} with the minus sign, and one mode decreases in mass, as in \eqq{eq.isospinchargedspectrum} with the plus sign. Note that the latter mode would be  massless  if we ignored the last term on the right-hand side in \eqq{eq.isospinchargedspectrum}. This would be the result for the energy of a macroscopic string stretched between the branes,  along the lines of \cite{Erdmenger:2006bg}. As in \cite{Erdmenger:2007vj}, the correction to this result given by the last term in \eqq{eq.isospinchargedspectrum} comes from the second-quantized dynamics associated to the endpoints of the string inherent in the D7-brane worldvolume action, in which these endpoints are described by fields instead of point-particles.

\subsection{Additional ungapped modes} \label{sec.moreungapped}
The massless modes associated to the broken SU(2)$_\Rsym$ symmetries can be thought of as the well known orientation zero modes of the instanton. An instanton in flat space has  additional  zero modes associated to the choice of the instanton centre. 
We will now show that these give rise to ungapped modes in our system. This may be surprising at first sight given the fact that the background geometry \eqq{eq.harmonicfunction} is not invariant under translations along the $y^i$-directions due to the preferred origin chosen by the harmonic function \eqq{eq.harmonicfunction}. Nevertheless, the exact cancellations due to supersymmetry guarantee that an instanton centred at any point in the $y^i$-directions is a solution. In other words, supersymmetry implies the existence of additional ungapped modes despite the fact that these are not Goldstone modes. 

The general instanton solution with an arbitrary centre breaks   SU(2)$_\lef$ symmetry, so it is convenient  to abandon the use of left-invariant forms in favor of the anti self-dual 't Hooft symbols, $\overline \eta_{ij}^{\hat a}$. These matrices connect the symmetries of $\mathbb{R}^4$ with those of SU(2)$_\fla$. We give here the form of the matrices of interest and list their properties in Appendix \ref{app.su2}:
\be\label{eq.thooftsymbols}
\overline \eta_{ij}^\pm  =  \frac{1}{\sqrt{2}} \left( \overline \eta_{ij}^1 \pm i \, \overline \eta_{ij}^2 \right) = \frac{1}{\sqrt{2}} \begin{pmatrix} 0 & 0 & \mp i & -1 \\ 0 & 0 & 1 & \mp i \\ \pm i & -1 & 0 & 0 \\ 1 & \pm i & 0 & 0  \end{pmatrix} \ , \quad
\overline \eta_{ij}^H  = \overline \eta_{ij}^3  = \begin{pmatrix} 0 & 1 & 0 & 0 \\ -1 & 0 & 0 & 0 \\ 0 & 0 & 0 & -1 \\ 0 & 0 & 1 & 0  \end{pmatrix} \ .
\ee
The ansatz \eqref{eq.ansatz} for the internal part of the gauge potential reads in this language
\be\label{eq.tHooftansatz}
A_\mt{mag} = \left( \frac{1}{\sqrt{2}} \overline \eta_{ij}^+ \otimes E_{-\alpha} + \frac{1}{\sqrt{2}} \overline \eta_{ij}^- \otimes E_{+\alpha} + \frac{1}{2} \overline \eta_{ij}^H \otimes H_\alpha \right) \d y^i \, \partial_j \log
\frac{1}{1 + a(y)} \,.
\ee
In this expression we have written  the SU(2)$_\fla$ generators in terms of ladder operators, $E_{\pm\alpha}$, and $H_\alpha$ (the Cartan subalgebra):
\be
E_{\pm \alpha} = \frac{1}{2} \left( \sigma^1 \pm i \sigma^2 \right) \ , \qquad H_\alpha =  \sigma^3 \,,
\ee
The subindex $\alpha$  is redundant for an SU(2) algebra but it will be important when analysing higher-rank algebras later. 
The self-duality condition in Eqn.~\eqref{eq.BPS} implies 
\be\label{eq.tHooftinstanton}
\Big[ 1 + a(y) \Big] \, \delta^{ij} \, \partial_i \partial_j  \, 
\frac{1}{1 + a(y)}
 = 0  \ .
\ee
On the other hand, we can perform an SU(2)$_\fla$ rotation to align the embedding along the direction of the Cartan. It is then a direct calculation to proof that, once a solution to \eqref{eq.tHooftinstanton} is found, the embedding and electric field are given by
\be\label{eq.HiggstHooft}
A_t=Z = \Mq \, \Big(1 + a(y) \Big) \, H_\alpha \ .
\ee
Eqn.~\eqq{ar} is a solution to \eqq{eq.tHooftinstanton} with spherical symmetry. A more general solution is given by 
\be\label{eq.offcentredinstanton}
\frac{1}{1 + a(y)} = 1 + \sum_{A=1}^k 
\frac{\Lambda_A^2 }{\delta_{ij} (y-y_A)^i (y-y_A)^j }  \ , 
\ee
corresponding to a sum of $k$ instantons with sizes $\Lambda_A$ and centred at $\vec y=\vec y_A$. The total instanton charge is $k$ and the different conserved charges such as $\Nr$, $\Ni$, $J$, $Q$, 
$\mathcal{E}$, etc. take the same form as in previous sections with the identification
\be
\label{combi}
\Lambda^2 = \sum_{A=1}^k \Lambda_A^2 \,.
\ee 
The solution with a single instanton takes the form
\be\label{single}
\frac{1}{1 + a(y)} = 1 + 
\frac{\Lambda^2 }{(\vec y-\vec y_0)^2 }  \,, 
\ee
and setting $\vec y_0=0$ we recover \eqq{ar}. As in \eqq{eq.goldstonescaling}, expanding to linear order in $\vec y_0$ around 
\mbox{$\vec y_0=0$} produces a normalisable change in the different fields that may be viewed as a fluctuation around the solution \eqq{ar}. These fluctuations break SU(2)$_\lef$ and are therefore not captured by \eqq{eq.Zfluc}-\eqq{eq.Afluc}. Nevertheless, they give rise to massless modes, since the energy density is independent of the value of $\vec y_0$. From the physical viewpoint, the freedom to place the instanton centres at arbitrary points reflects the no-force condition between the D3-branes in the D3-D7 system when supersymmetry is preserved. On the gauge theory side the positions of the centres correspond to expectation values of higher-dimension operators that parameterise the Higgs branch and are bilinear in the fundamental fields, in analogy with the case of the Coulomb branch of $\mathcal{N}=4$ SYM discussed  in \cite{Klebanov:1999tb}. 

Note that only the combination \eqq{combi} is fixed by the conserved charges of the solution. Therefore, for multi-instanton solutions, combinations of the instanton sizes $\Lambda_A$ giving rise to the same total $\Lambda^2$ in \eqq{combi} also generate additional massless modes.

\section{Higher-rank flavor groups ($\nf > 2$) 
\label{higher}}
In this section we will generalise our previous construction to the case in which the number of flavors is $\nf > 2$.  The strategy is to write the generators of a semisimple Lie algebra in canonical form in terms of Cartans and ladder operators in order to sequentially solve for the self-dual part of the field strength and, once this solution is obtained, for the embedding profile. The SU(2)$_\fla$ instanton presented in previous sections can be used as a building block to find solutions for higher-rank gauge groups, in particular for SU($\nf$). In this subsection we restrict the discussion to SU(3)$_\fla$ since this case already illustrates all the new aspects while remaining concrete enough. We will nevertheless use a notation appropriate to make contact with the  generic case, which  is discussed in Appendix \ref{app.su2}. One of these new   features is the existence of new embeddings of the flavor branes.

The starting point are the Gell-Mann matrices, which can be combined in three groups of ladder operators
\be
E_{\pm\alpha} = \frac{1}{2} \left( \lambda_1 \pm i \lambda_2 \right) \ , \qquad E_{\pm\beta} = \frac{1}{2} \left( \lambda_6 \pm i\lambda_7 \right) \ , \qquad E_{\pm\gamma} = \frac{1}{2} \left( \lambda_4 \pm i \lambda_5 \right) \,,
\ee
together with the Cartan generators
\be\label{eq.SU(3)cartans}
H_1 = \frac{1}{\sqrt{2}} \, \lambda_3 \ , \qquad H_2 = \frac{1}{\sqrt{2}} \, \lambda_8 \ ,
\ee
which, by definition, form an Abelian subalgebra. Here $\hat\alpha=(\alpha,\beta,\gamma)$ refers to the three roots of SU(3) 
\be
\alpha = \left(\sqrt{2},0\right) \ , \qquad \beta = \left(-\frac{1}{\sqrt{2}},\sqrt{\frac{3}{2}}\right) \ , \qquad \gamma = \alpha+\beta = \left(\frac{1}{\sqrt{2}},\sqrt{\frac{3}{2}}\right) \ .
\ee
Note that each root is a two-vector. We will collectively denote the components of these three vectors as $\hat \alpha_{\hat \imath}$ with 
$\hat \imath=1,2$. Similarly, we will collectively denote the Cartans \eqq{eq.SU(3)cartans} as $H_{\hat \imath}$. The ladder operators are eigenvectors of $[H_{\hat \imath},\cdot]$ with
\be
[H_{\hat \imath}, E_{\pm \hat\alpha}] = \pm \hat \alpha_{\hat \imath} E_{\pm \hat\alpha} \,.
\ee
It is convenient to define the following linear combinations of the Cartan generators
\be
H_{\hat \alpha} \equiv \left[ E_{+\hat \alpha}, E_{-\hat \alpha} \right] = \hat \alpha_{\hat\imath} H_{\hat\imath}
\ee
because then each triplet $\{E_{\pm \hat\alpha}, H_{\hat\alpha}\}$ associated to each root vector spans an SU(2)  subalgebra (in which $H_{\hat\alpha}$ is not necessarily canonically normalised). Each of these subalgebras can then be used to construct an SU(2)$_\fla$ instanton embedded inside SU(3)$_\fla$. 

With this material in place we focus on solutions that can be obtained via the 't Hooft ansatz. Without loss of generality  we choose to align the magnetic part of the gauge potential along the first root, $\alpha$. This can always be achieved with an SU(3)$_\fla$ rotation. We therefore write the magnetic part of the gauge potential as in \eqref{eq.tHooftansatz}, so the self-duality condition of the SU(2)$_\fla$ instanton leads again to Eqn.~\eqref{eq.tHooftinstanton} for the function $a(y)$. Since we have fixed the  orientation of the instanton we are no longer free to choose the orientation of the scalar field $Z$. Therefore, in the general solution $Z$ should be a linear combination of all the real generators of SU(3)$_\fla$ which cannot be rotated away while keeping the orientation of the instanton fixed. These  are 
\be
H_\alpha \sac H_\beta  \sac 
 E_{+\beta} + E_{-\beta} \sac  E_{+\gamma} + E_{-\gamma} \,.
 \ee
We have not included $H_\gamma = H_\alpha+H_\beta$ because this is not linearly independent (see Appendix \ref{app.su2}), and we have not included $E_{+\alpha} + E_{-\alpha}$ because this can be rotated away while keeping the instanton orientation invariant. A direct embedding in SU(3)$_\fla$ of the SU(2)$_\fla$ solution of previous sections would yield
\be
\label{directimport}
A_t=Z = M_\alpha \, \Big[ 1 + a(y) \Big] H_\alpha \,,
\ee
where we have relabelled $\Mq$ as $M_\alpha$. Now, inside SU(3)$_\fla$, we have two qualitatively distinct possibilities to augment this solution. We can either add terms proportional to the second Cartan generator, or we can add terms proportional to the real combinations of the ladder operators. The result of adding a linear combination of these possibilities is the linear combination of the results for each individual possibility, so we will consider them separately. 

\begin{figure}
    \centering
    \begin{subfigure}[t]{0.47\textwidth}
        \centering
        \includegraphics[width=0.85\linewidth]{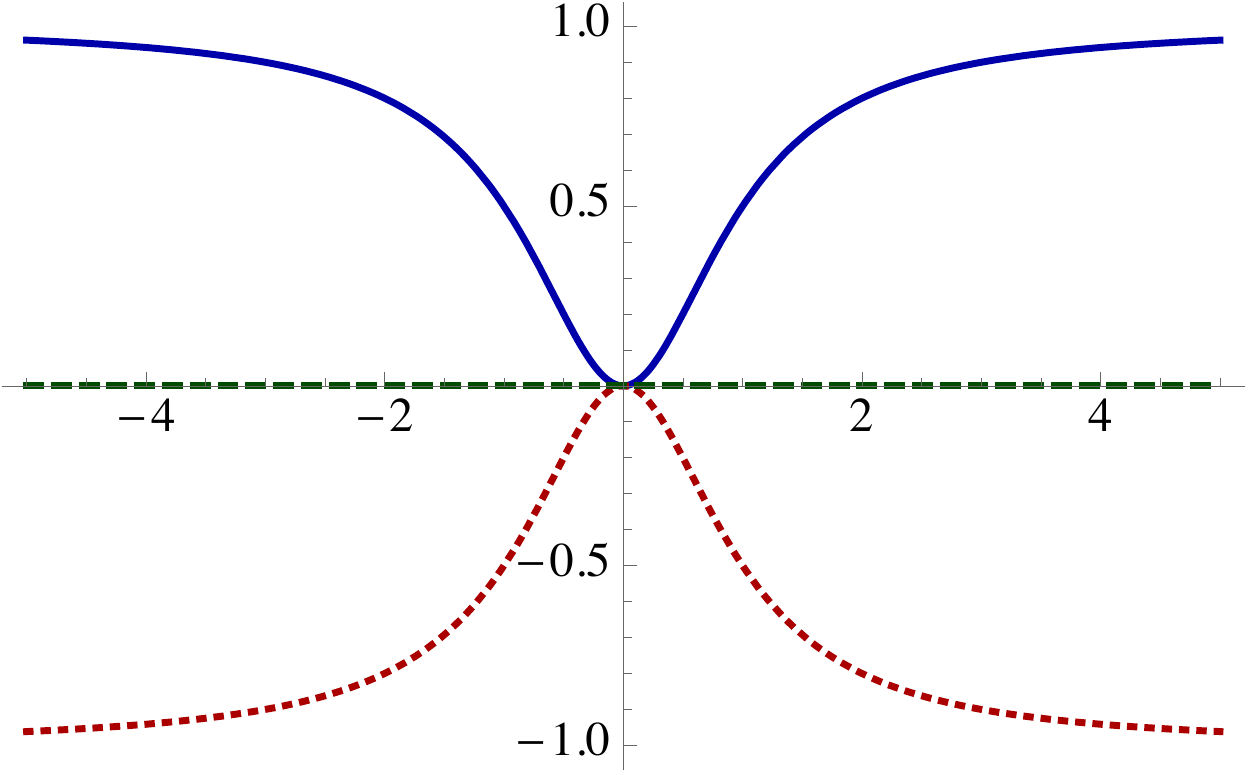} 
        \caption{$M_\beta=0$} \label{Zcartan0}
    \end{subfigure}
    \begin{subfigure}[t]{0.47\textwidth}
        \centering
        \includegraphics[width=0.85\linewidth]{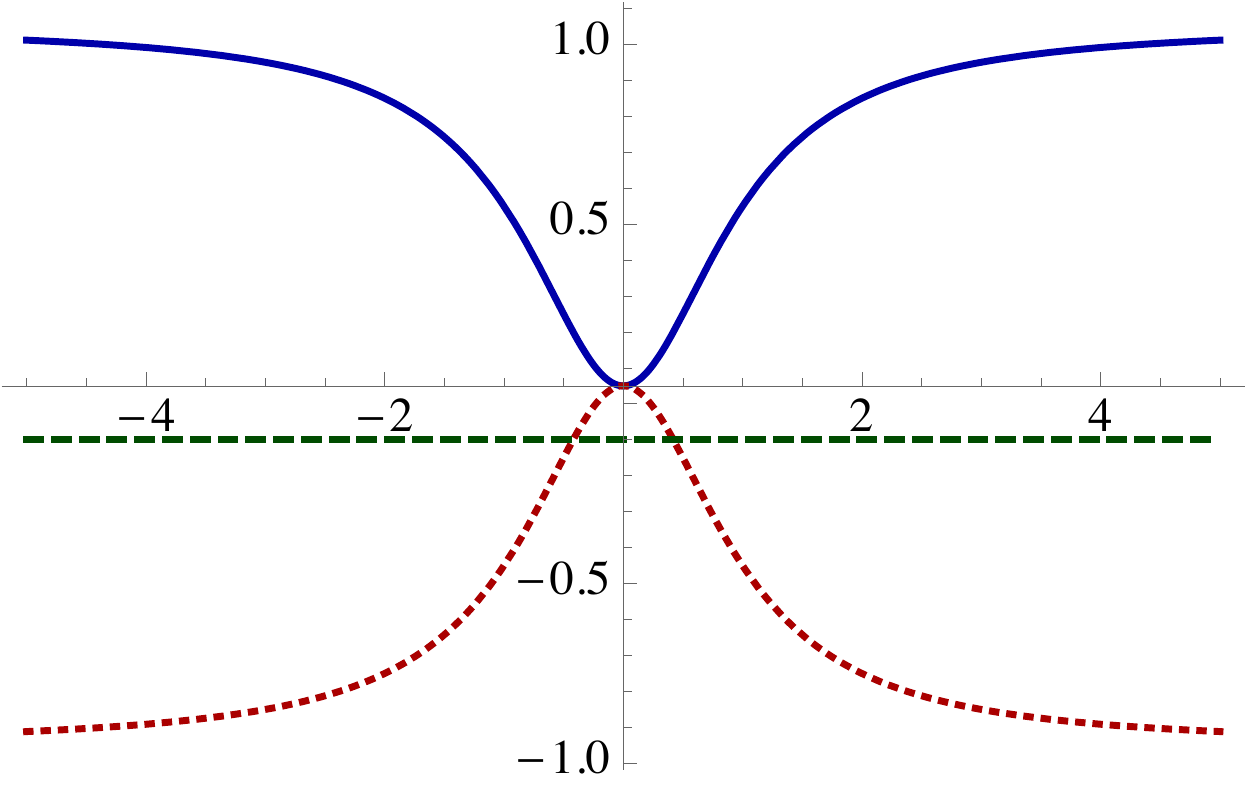} 
        \caption{$M_\beta=1/10$} \label{Zcartantenth}
    \end{subfigure}

\vspace{.5cm}
    \begin{subfigure}[t]{0.47\textwidth}
        \centering
        \includegraphics[width=0.85\linewidth]{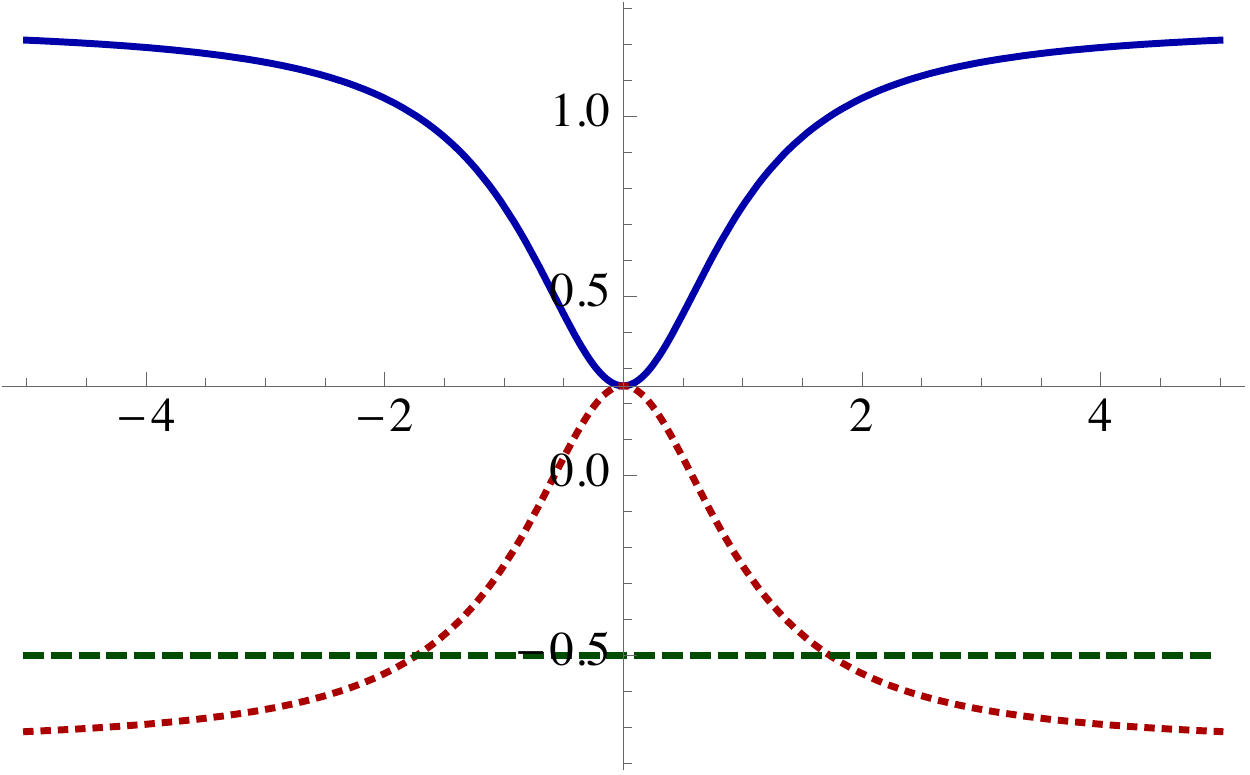} 
        \caption{$M_\beta=1/2$} \label{Zcartanhalf}
    \end{subfigure}
    \begin{subfigure}[t]{0.47\textwidth}
      \centering
      \includegraphics[width=0.85\linewidth]{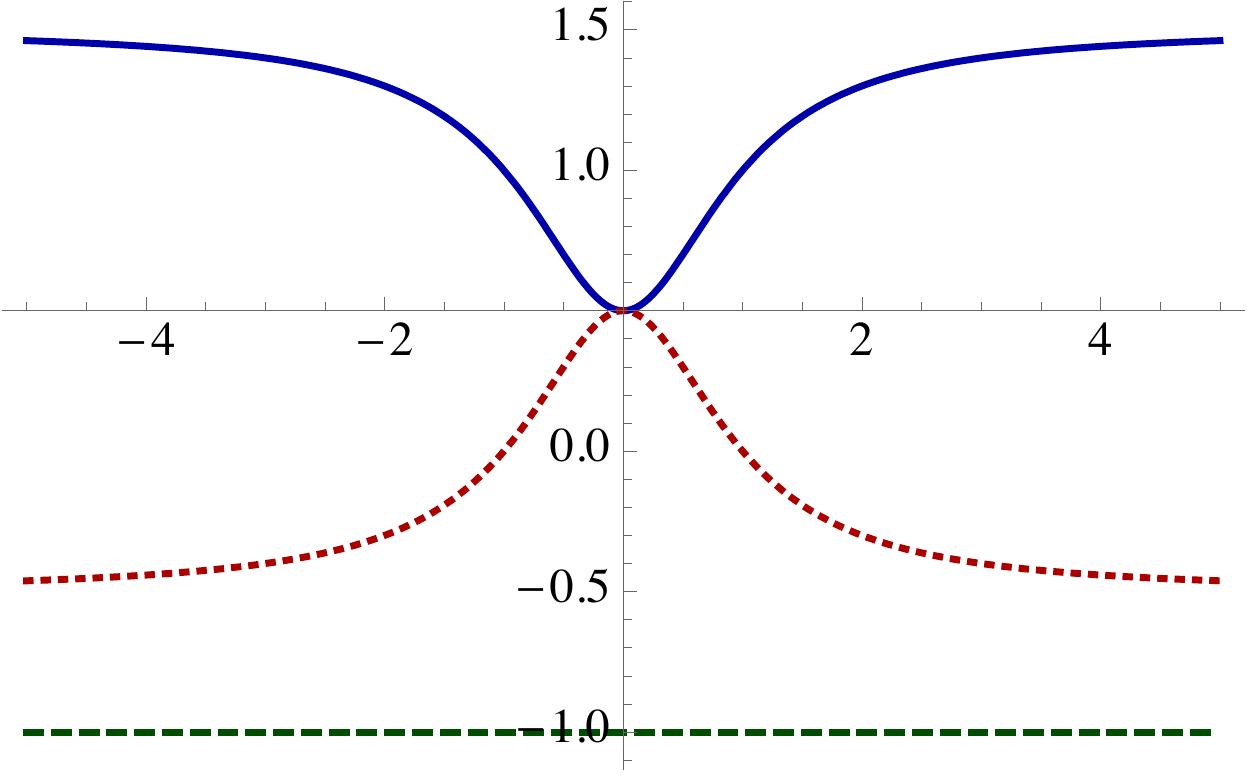} 
      \caption{$M_\beta=1$} \label{Zcartan1}
    \end{subfigure}
    \caption{\small Brane embeddings of the form \eqref{cartanemb} for different values of $M_\beta$. In all the plots we have fixed $M_\alpha=1$.}
    \label{fig.Zcartan}
\end{figure}

For the first option, the boundary conditions at the boundary and the regularity requirement at the origin lead to the solution (see Appendix \ref{app.su2})
\be\label{cartanemb}
A_t=Z = \left( M_\alpha \, \Big[ 1 + a(y) \Big] + \frac{M_\beta}{2} \right) H_\alpha + M_\beta H_\beta \ ,
\ee
where $M_\beta$ is now a second possible mass term. Since the scalar field is still diagonal, the geometric interpretation follows straightforwardly from the three entries of the $Z$ matrix:   
 \be
Z_{\pm} = \frac{M_\beta}{2} \pm M_\alpha \, \Big[ 1 + a(y) \Big] \ , \qquad Z_3 = - M_\beta \,.
\ee
These determine the embedding profiles of the three D7-branes in the $z^1$-direction. For a single instanton centred at the origin of AdS$_5$ the radial dependence is given by Eqn.~\eqref{eq.solution} and the profiles are shown in Fig.~\ref{fig.Zcartan}. 
If the embedding is purely along $H_\alpha$, that is, if $M_\beta=0$ as in the direct import of the SU(2)$_\fla$ solution \eqq{directimport}, then two of the branes have the same embedding as in previous sections and the third one describes exactly massless quarks. A non-zero value of $M_\beta$ produces a relative displacement of the branes while keeping their centre of mass at $z^1=0$, since the SU(3)$_\fla$ matrices are traceless. In particular, for non-zero $M_\beta$ the branes no longer reach the origin of AdS at $r=0$, as seen in the figure. Moreover, for intermediate values of $M_\beta$ the brane profiles can cross at non-zero $r$, meaning that the hierarchy of running quark masses in the IR may be inverted with respect to that in the UV. For sufficiently large values of $M_\beta$ this inversion disappears.  The non-compact part of the induced metric on the D7-branes is always ${\rm AdS}_5$  in the UV. If $M_\beta=0$ then $Z_{\pm} \propto r^2$ near $r=0$, as in previous sections. In this case the IR  geometry is again  AdS$_5$ with the same radius as in the UV, as in \Sec{IRlimit}. If instead $M_\beta\ne0$ then the IR metric is flat space, as for the usual massive embedding in the absence of instanton and electric field. 

\begin{figure}
    \centering
    \begin{subfigure}[t]{0.47\textwidth}
        \centering
        \includegraphics[width=0.85\linewidth]{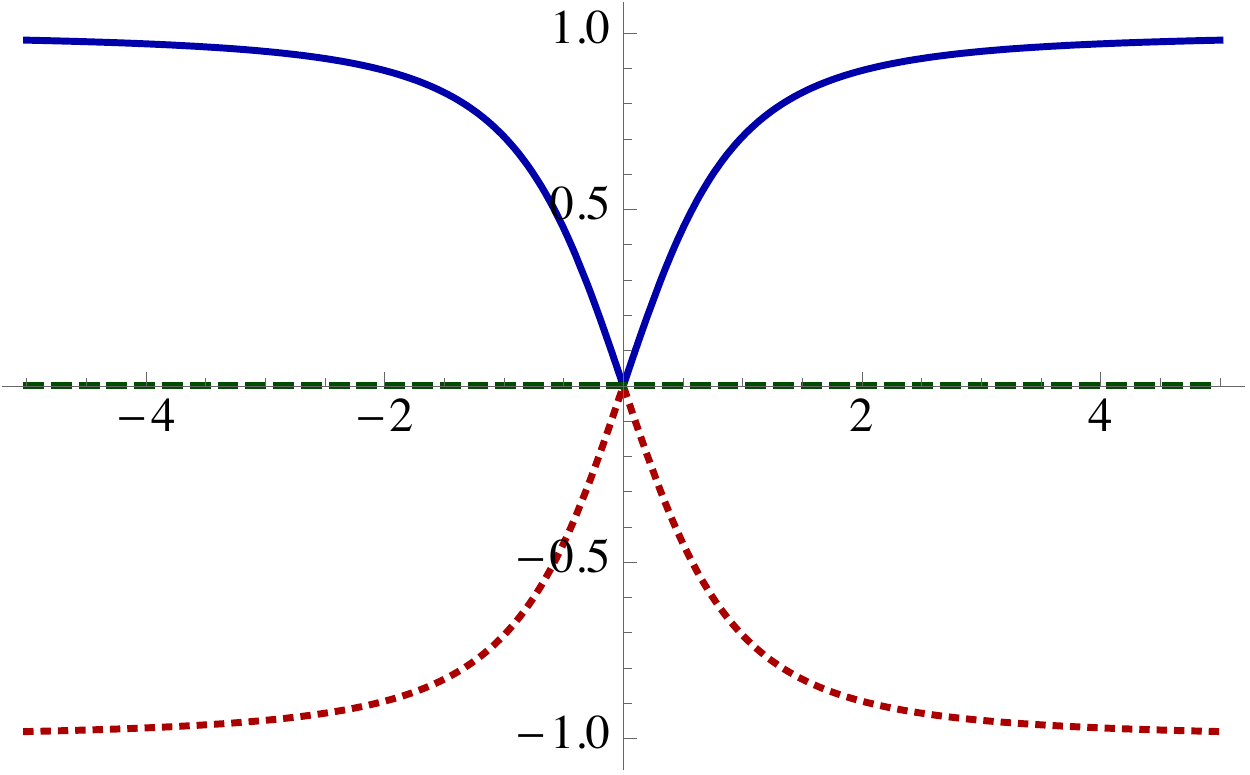} 
        \caption{$M_\alpha=0$} \label{Zbeta0}
    \end{subfigure}
    \begin{subfigure}[t]{0.47\textwidth}
        \centering
        \includegraphics[width=0.85\linewidth]{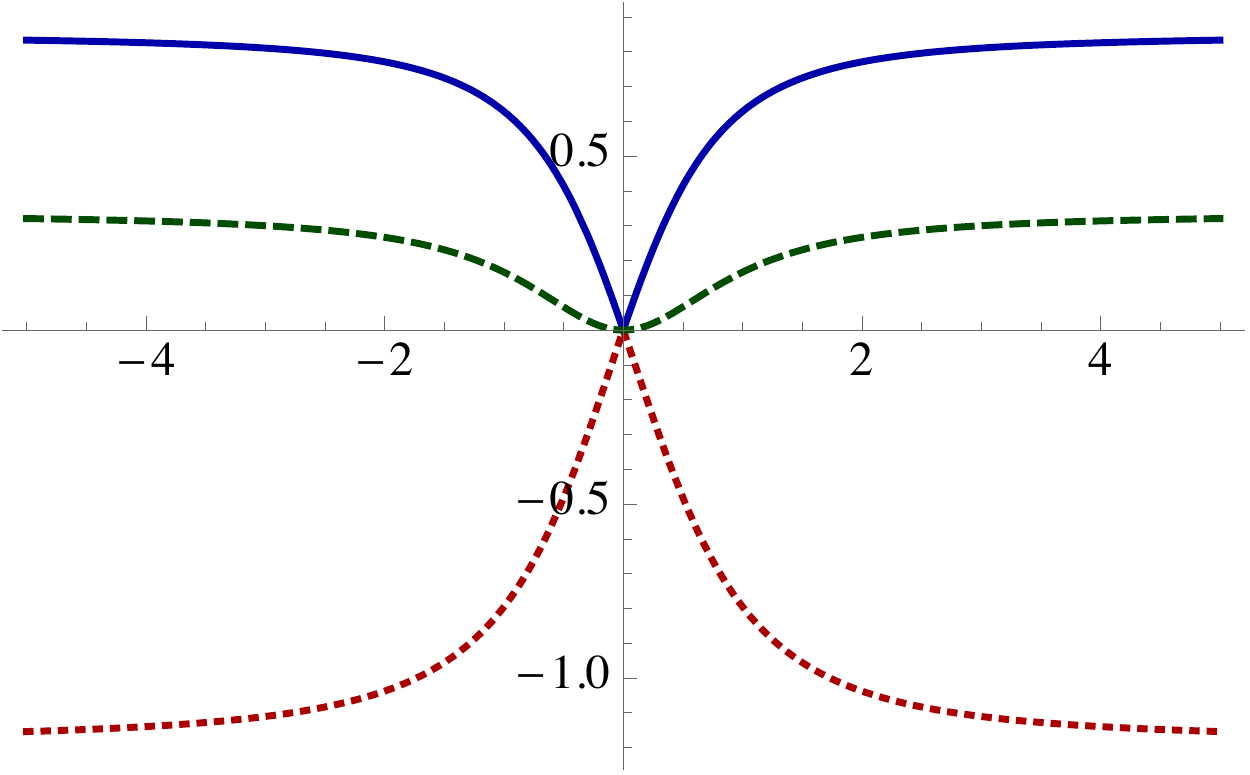} 
        \caption{$M_\alpha=1/3$} \label{Zbetathird}
    \end{subfigure}
\vspace{.5cm}
    \begin{subfigure}[t]{0.47\textwidth}
        \centering
        \includegraphics[width=0.85\linewidth]{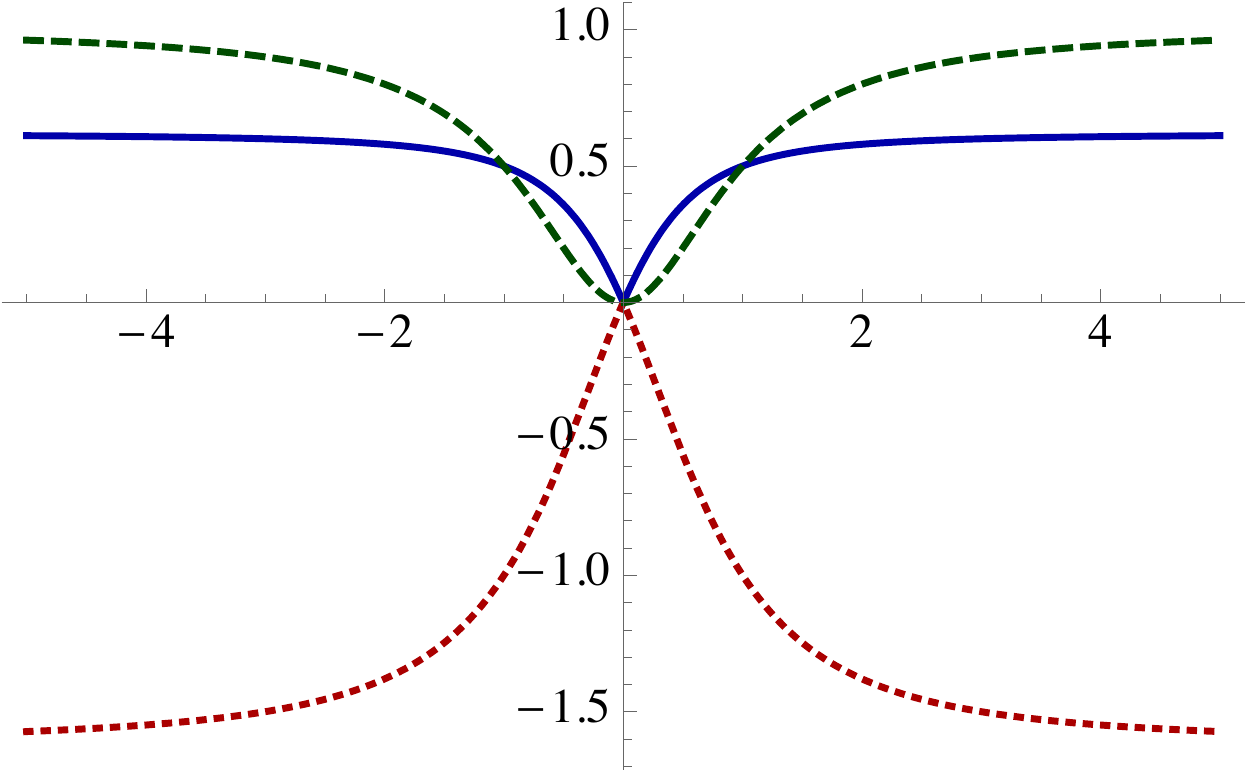} 
        \caption{$M_\alpha=1$} \label{Zbeta1}
    \end{subfigure}
    \begin{subfigure}[t]{0.47\textwidth}
      \centering
      \includegraphics[width=0.85\linewidth]{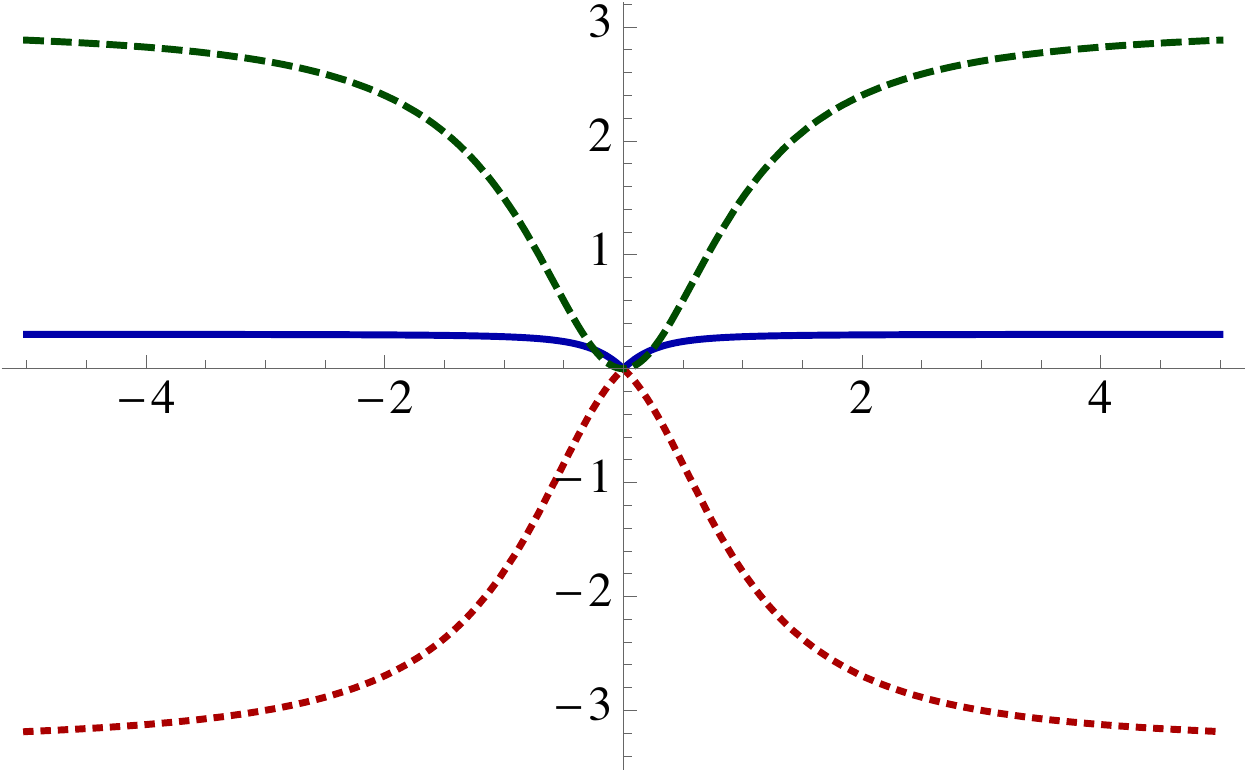} 
      \caption{$M_\alpha=3$} \label{Zbeta3}
    \end{subfigure}
    \caption{\small Brane embeddings of the form \eqref{secondsol} for different values of $M_\alpha$. In all the plots we have fixed $M_\beta=1$.}    \label{fig.Zbeta}
\end{figure}

For the second possibility it suffices to consider adding a single combination of the form $E_{+\beta} + E_{-\beta}$. In this case the regular solution is 
\be
\label{secondsol}
A_t=Z =  M_\alpha \Big[ 1 + a(y) \Big] \, H_\alpha + M_\beta \sqrt{ 1 + a(y) } \,\Big( E_{+\beta} + E_{-\beta} \Big) \ .
\ee
These embeddings are not diagonal in this basis, but this is simply due to the fact that we fixed the instanton orientation. Having found the solution \eqq{secondsol} we could simply apply an SU(3)$_\fla$ rotation to bring $Z$ to a diagonal form at the cost, of course, of having an instanton no longer contained in the  SU(2) algebra generated by $\{E_{\pm \hat\alpha}, H_{\hat\alpha}\}$. From the viewpoint of the D7-branes this rotation is simply a gauge transformation. The result would be a diagonal $Z$ with 
entries equal to its  eigenvalues 
\be
Z_\pm = \frac{\sqrt{ 1 + a } }{2} \Big[ M_\alpha \sqrt{ 1 + a }  \pm \sqrt{M_\alpha^2 ( 1 + a )  + 4 M_\beta^2} \, \Big] \ , \qquad Z_3 = M_\alpha  ( 1 + a ) \,.
\ee
These are plotted in Fig.~\ref{fig.Zbeta} for the instanton in Eqn.~\eqref{eq.solution} and, as above, they can be interpreted as the positions of the branes. 
Note that there is no ambiguity in this interpretation because we are only allowing one transverse scalar $Z^1$ to be excited. Generically, in the presence of two scalars $Z^1, Z^2$ it would not be possible to diagonalise both of them simultaneously. As in the previous case, there is a running mass hierarchy inversion in some range of $M_\beta/M_\alpha$, but the branes always reach the origin in the deep IR. Note that, if $M_\beta$ is not zero, then the leading behavior of $Z_\pm$ near the origin is 
\be
\label{same}
Z_\pm \sim M_\beta \sqrt{1+a} \sim r \,,
\ee
and exactly the same for the corresponding components of $A_t=Z$. As in previous cases, the electric field contribution to the energy density, $\tr\delta^{ij}E_iE_j$, vanishes at $r=0$ and its total contribution is given by 
\eqq{total}. The behavior \eqq{same} is the same as in \cite{Karch:2007br} except for the fact that the supersymmetric limit $A_t=Z$ can not be achieved in the context of \cite{Karch:2007br} due to the absence of the instanton. As before, the induced metric is ${\rm AdS}_5\times{\rm S}^3$ both in the UV and IR, but the softening in the vanishing of $Z$, as dictated by \eqref{same}, produces a renormalization in the radius of the sphere along the RG flow
\be
\frac{L_\mt{UV}^2}{L_\mt{IR}^2} = 1 + \frac{M_\beta^2}{\Lambda^2}\,.
\ee

\section{Supertubes}
\label{super}

In the presence of angular momentum, a collection of D0-branes and fundamental strings is known to ``blow up'' into a configuration known as a D2-brane supertube \cite{Mateos:2001qs}, namely a tubular D2-brane with the same charges as the original system. The supergravity description of two-charge supertubes was found in \cite{Emparan:2001ux} and the generalisation to three charges was presented in \cite{Elvang:2004rt,Elvang:2004ds,Bena:2004de,Gauntlett:2004wh,Gauntlett:2004qy}.  The cross-section of the tube is completely arbitrary and contractible, hence there is no net D2-brane charge. Such a tubular D2-brane can be suspended between D4-branes. The entire configuration preserves supersymmetry and appears on the U(2) non-Abelian theory on the worldvolume of the D4-branes as a dyonic instanton configuration \cite{Kim:2003gj}. T-dualising four times we conclude that, in the presence of angular momentum,  our system of D3-branes and fundamental strings  should blow up into a D5-brane supertube suspended  between the D7-branes. The defining feature of this configuration should be that the D7-branes meet each other at a curve in $\mathbb{R}^4$. Moreover, for these configurations the angular momentum is not necessarily self-dual. The reason that we have not encountered these features in the  configurations of previous sections  is that we did not consider the most general dyonic instanton solution. Here we will import the results of \cite{Kim:2003gj} for the $k=2$ case to illustrate the fact that, in the general solution, the D7-branes indeed meet at a curve. The solutions of previous sections can then be obtained as a  limit in which the cross-section of the supertube collapses to a set of isolated points. Our goal is not to review the details of \cite{Kim:2003gj}, to which we refer the interested reader, but merely to illustrate the main points. 

The starting point in \cite{Kim:2003gj} is the most general  solution to \eqref{eq.tHooftinstanton}, the so-called Jackiw--Nohl--Rebbi instanton \cite{Jackiw:1976fs}:
\be\label{eq.JNR}
\frac{1}{1+a} = \sum_{A=0}^k \frac{\Lambda_A^2}{\delta_{ij}(y-y_A)^i (y-y_A)^j} \ .
\ee
The difference between this solution and \eqref{eq.offcentredinstanton} is that there is no ``1'' on the right-hand side and that there are $k+1$ ``centres'' despite the fact that the instanton number is $k$. Thus compared to \eqref{eq.offcentredinstanton}, naively this solution depends on one new scale and on four new centre coordinates.  However, for $k=1,2$ there are additional gauge equivalences that reduce the number of these new parameters to 2 and 4, respectively \cite{Jackiw:1976fs}. 
Solutions of the 't Hooft type can be recovered in two different ways:
\begin{itemize}
\item for generic $k$ one can obtain \eqref{eq.offcentredinstanton} by taking the limit $\Lambda_0\to\infty$ with $|y_0|^2/\Lambda_0^2=1$, 
\item for $k=1$ both solutions \eqref{eq.offcentredinstanton} and \eqref{eq.JNR} are related by a gauge transformation in such a way that the parameters in both solutions are related through 
\be
y_c=\frac{\Lambda_1^2 y_0 +\Lambda_0^2 y_1}{\Lambda_0^2 + \Lambda_1^2} \ , \qquad 
\Lambda_c^2 = \Lambda_0^2 \Lambda_1^2 \frac{|y_1-y_0|^2}{(\Lambda_0^2 + \Lambda_1^2)^2} \,,
\ee
where the parameters in \eqref{eq.offcentredinstanton} are on the left-hand side  and are denoted with a \mbox{subindex ``c''}.

\end{itemize}

The absence of a ``1'' on the right-hand side of \eqq{eq.JNR} changes the asymptotics of the gauge potential and makes finding a regular solution for the scalar field a difficult task. The procedure to follow is described in detail in \cite{Kim:2003gj,Choi:2007mk} and we refer the reader to those references  for details on the construction. The final result is that such solutions exist, but the field $Z=Z_{\hat a} \, \sigma^{\hat a}$ is not necessarily aligned in SU(2)$_\fla$ with its asymptotic value $\Mq \sigma^3$. Moreover, if $k\geq2$ then generically the zeros of $Z$ lie on a curve instead of on a set of isolated points. An example is given in Sec.~4.2 of \cite{Kim:2003gj} for $k=2$, whose main result we now review. 

\begin{figure}[t]
\begin{center}
\includegraphics[width=.65\textwidth]{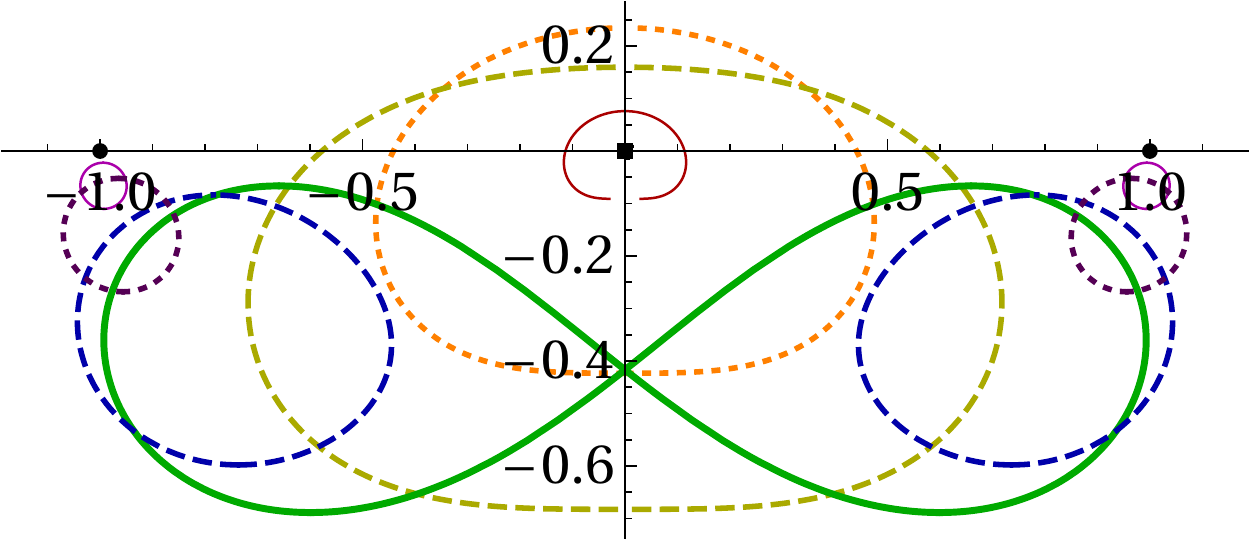} 
\put(-148,130){\mbox{\large $y^4$}}
\put(2,88){\mbox{\large $y^3$}}
\end{center}
\vspace{1mm}
\caption{\small Cross-section of the supertube at $y^1=y^2=0$ for different values of $\Lambda_0$. For $\Lambda_0=0$ the supertube collapses to  the $k=1$ instanton configuration centred at the origin 
(denoted by a black square) that we have studied in detail in this paper. 
For $0< \Lambda_0\approx 2.285$ the zeros of the scalar field lie on a simply connected, non-self-intersecting curve (a  deformed ellipse). This is illustrated by the values 
 $\Lambda_0=1/10$ (thin, red, continuous curve), $\Lambda_0=1/2$ (orange, dotted curve) and $\Lambda_0=1$ (yellow, dashed curve).  
 At $\Lambda_0\approx 2.285$ (thick, green, continuous curve) this curve pinches off. For $2.285 \lesssim \Lambda_0 < \infty$ the zeros of the scalar field lie on two disconnected curves, as illustrated by the values 
 $\Lambda_0=3$ (blue, dashed curve), $\Lambda_0=8$ (purple, dotted curve), and $\Lambda_0=20$ (thin, magenta, continuous curve). 
For $\Lambda_0\to \infty$ these two curves collapse to two points denoted by  the black circles on the $y^3$ axis. 
\label{fig.supertubes}
}
\end{figure}

For simplicity consider the three centres to lie within the 34-plane of $\mathbb{R}^4$: 
\be
y_0 = \left( 0,0,0,-\Lambda_0 \right) \ , \quad y_1 = \left( 0,0, \Lambda \frac{\Lambda_1}{\Lambda_2} ,0 \right) \ , \quad y_2 = \left( 0,0, - \Lambda \frac{\Lambda_2}{\Lambda_1} ,0 \right) \,.
\ee
For $\Lambda_0=0$ we effectively recover the $k=1$ case, which describes an instanton centred at $y_c= 0$ with size $\Lambda_c=\Lambda$, as has been considered throughout this paper. In the opposite limit, $\Lambda_0\to\infty$, one recovers the two instanton solution of \eqref{eq.offcentredinstanton} with centres at $y_1$ and $y_2$. Thus, in these two limits the zeros of the scalar field correspond to isolated points. For any other value of $\Lambda_0$ it can be shown that the zeros lie on the curve in $\mathbb{R}^4$ determined by the solution to the equation \cite{Kim:2003gj}
\be
\label{soso}
\bal
\sum_{A=0}^2 & \Big[ \, \Lambda_A^4 |y-y_{A+1}|^2 |y-y_{A+2}|^2 + 2 \Lambda_{A+1}^2 \Lambda_{A+2}^2 |y-y_{A}|^2 (y-y_{A+1})\cdot (y-y_{A+2}) \\
& \quad + {\cal{P}}'\, |y-y_{A}|^2 \, (y-y_{A+1}) \,\overline \eta^3 (y-y_{A+2}) \Big] = 0 \,.
\eal
\ee
In this expression the index is understood to be cyclic, i.e. $y_3=y_0$ and $y_4=y_1$ and similarly for the $\Lambda_A$'s, $\overline \eta^3$ is the anti-self-dual 't Hooft matrix \eqref{eq.thooftsymbols},  the constant ${\cal{P}}'$ is given by 
\be
{\cal{P}}' =  4\, \frac{ y_0 \, \eta^3 y_1 + y_1 \, \eta^3 y_2 + y_2 \, \eta^3 y_0}{ \Lambda_0^{-2} \Lambda_1^{-2} |y_0-y_1|^2 + \Lambda_1^{-2} \Lambda_2^{-2} |y_1-y_2|^2 + \Lambda_2^{-2} \Lambda_0^{-2} |y_2-y_0|^2 } \,,
\ee
and $\eta^3$ is the self-dual 't Hooft matrix that can be obtained from $\overline \eta^3$ by flipping  the signs of the last row and of the last column. By symmetry the zeroes lie in the $y^1=y^2=0$ plane inside $\mathbb{R}^4$, so \eqq{soso} is one equation for the two non-zero components of $y$ in the 34-plane. The solution is generically a curve in this plane. In \fig{fig.supertubes} we show several examples where we set 
$\Lambda=\Lambda_1=\Lambda_2=1$ for simplicity and we vary 
 $\Lambda_0$. As can be seen in the figure, the single-instanton solution considered in this paper corresponds to a collapsed supertube.

\section{Conclusions\label{sec.conclusions}}

We have considered a configuration of two D7-branes in the background created by a stack of $\nc$ coincident D3-branes, holographically dual to probing $d=4$, $\mathcal{N}=4$, SU($\nc$) super Yang--Mills with a fundamental  $\mbox{SU(2)}_\fla$-doublet, $\mathcal{N}=2$  hypermultiplet at finite isospin and R-charge densities. On a technical note, our model requires dealing with the non-Abelian version of the Dirac--Born--Infeld action, which is known to be an incomplete account of the dynamics of flavor with a global U($\nf$) symmetry group \cite{Tseytlin:1997csa,Myers:1999ps}. However, as we have argued, the physics of supersymmetric configurations can be captured exactly  by a Yang--Mills--Higgs action  \cite{Hashimoto:1997px,Bak:1998xp}. 

The ground state preserves ${\cal N}=1$ supersymmetry and can be used to construct generalisations to $\nf\ll\nc$ D7-branes. This solution corresponds holographically to a state on the field theory presenting spontaneous breaking of the (non-Abelian) global and local symmetries. The  broken global symmetries are linked to superfluidity in the system. 
The breaking of the local symmetries is encoded in the dissolution of D3-brane charge on the D7-branes' worldvolume, with the associated energy scale $\Lambda$, determined by the asymptotic charges of the system.

The solution we have presented is regular everywhere. In the IR the induced metric on the D7-branes is $\mbox{AdS}_5 \times \mbox{S}^3$, with active irrelevant deformations due to the non-trivial configuration of the instanton there. Although not obvious from the AdS$_5$ factor in the metric, Lorentz symmetry is broken by the presence of non-trivial sources of Lorentz-breaking operators. This can be seen also from the presence of non-relativistic massless excitations in the spectrum.  Backreaction of the D7-branes could be incorporated  along the lines of \cite{Faedo:2016jbd}. In that reference we considered a configuration with an instanton but no isospin density, and studied the backreaction on the metric.  In \cite{Faedo:2016jbd}  Lorentz invariance is maintained but the presence of the D7-brane sources changes the metric at all energy scales. Furthermore, the breaking of the SU($\nc$) group is seen explicitly in the running of the $F_5$ flux, as explained in \Sec{sec.symmetrypatterns}.

We have found a set of massive quasiparticle excitations in the $\Mq\ll \Lambda$ regime of parameters, whose lifetimes decrease when $\Mq\sim \Lambda$. On top of these there is a set of massless excitations with non-relativistic dispersion relations, the Goldstone modes associated to the breaking of the global symmetries, as well as light particles with a mass gap proportional to the explicit scale $\Mq$ whenever $\Mq\ll \Lambda$. 
Interestingly, there is also a set of ungapped modes that are not associated to the breaking of any global symmetries but to exact moduli implied by supersymmetry. 

In the presence of angular momentum, it is well known that Dp-branes and fundamental strings get ``blown up'' into a D(p+2)-supertube \cite{Mateos:2001qs}. Importing the results from \cite{Kim:2003gj} we have shown that this is realised in our setup. As in \cite{Kim:2003gj} we have focused on the case of instanton number $k=2$ but more general results are possible. In the $k=2$ case we have shown that if the R-charges  \eqq{through2} are not equal, $n_1\neq n_2$, then the fundamental strings and the D3-branes dissolved inside the D7-branes get blown into a D5-brane supertube with non-self dual angular momentum suspended between the D7-branes. From the viewpoint of the D7-branes this manifests itself in the fact that they meet each other at a curve instead of at isolated points. If the charges are fine-tuned to be equal, so that the angular momentum becomes self-dual, then the cross-section of the supertube collapses to a set of isolated points.

In the future we expect to combine the results and ideas presented in \cite{Faedo:2018fjw} and in the present paper with our previous constructions of backreacted solutions of the charged D3/D7 system \cite{Bigazzi:2011it,Faedo:2017aoe} in order to find the holographic description of a color superconductor at non-zero baryon density. It could also be of interest to combine the present configurations with the finite temperature setups of \cite{Bigazzi:2009bk,Faedo:2016cih} to study the phase transition between the superconducting and the normal phases.

\section*{Acknowledgements}

We thank Andrea Amoretti and Carlos Hoyos for discussions. 
AF and DM are supported by grants FPA2016-76005-C2-1-P, FPA2016-76005-C2-2-P, 2014-SGR-104, 2014-SGR-1474, SGR-2017-754 and MDM-2014-0369. CP is supported by the STFC Consolidated Grant ST/P000371/1. JT is supported  by the European Research Council, grant no. 725369.

\appendix
\setcounter{tocdepth}{1}

\section{Proof of the BPS equations for the dyonic instanton\label{app.naDBI2HYM}}

In this appendix we demonstrate that the dyonic instanton of \cite{Lambert:1999ua} solves the equations of motion coming from the non-Abelian Dirac--Born--Infeld (DBI) action of \cite{Tseytlin:1997csa,Myers:1999ps}, not only in flat space \cite{Zamaklar:2000tc}, but also in the background of a stack of $\nc$ D$p$-branes (in the main text we set $p=3$).
Consider the action of a set of $\nf$ D$(p+4)$-branes, whose pull-back metric induced by the color D$p$-branes' geometry is
\be\label{Dp}
\d s^2 = H^{-\frac{1}{2}} \eta_{\mu\nu} \d x^\mu \d x^\nu + H^{\frac{1}{2}}\left(\delta_{ij} \, \d y^i \d y^j + \delta_{\alpha\beta}\, \d z^\alpha \d z^\beta \right) \ ,
\ee
where $\mu=0,\cdots, p$, $i=1,\cdots, 4$ and $\alpha=1, \cdots, 5-p$. The warp factor is an harmonic function in $\mathbb{R}^{9-p}$
\be\label{H}
H = L^{\frac{7-p}{2}}\left( \delta_{ij}\, y^i  y^j + \delta_{\alpha\beta}\, z^\alpha  z^\beta \right)^{\frac{p-7}{2}}\ .
\ee
This is supported by a non-trivial $\left(p+1\right)$-form together with a dilaton:
\be\label{Cp1}
C_{p+1} = q \, H^{-\frac{p+1}{4}}\d x^{1,p}\ , \qquad e^{\Phi} = H^{\frac{3-p}{4}} \ , 
\ee
with $q^2=1$ a sign distinguishing branes from anti-branes. The stack of flavor branes is extended in the $x$-directions as well as the $y$ ones describing $\mathbb{R}^4$. The $5-p$ remaining flat directions, parameterised by $z$, are transverse to both kinds of branes. There is also a U($\nf$) gauge field $A$, with field strength $F$, living in this set of branes.

The non-Abelian DBI action  in \cite{Tseytlin:1997csa,Myers:1999ps} is rather complicated, so we present it already adapted to our system. First, by SO(5-$p$) symmetry we align the transverse scalars in the $Z^{\alpha\neq1}=0$ direction and denote the remaining one $Z^1=Z$. In this case the scalar potential in \eqref{wvaction} is trivial and, in the background of the color branes, the kinetic action reduces to
\be
S_{DBI}  = -T_{p+4}\int\d^{p+1} x\d^4 y \,e^{-\Phi}\,{\rm Str} \sqrt{-\det\left( g_{MN} + {\cal P}[H]^{\frac{1}{2}}  D_{(M}Z D_{N)}Z + F_{MN}\right)} \ ,
\ee
where $M$ and $N$ are curved indices in the $(p+5)$ directions of the flavor branes. In the following we name the matrix inside the determinant $\Gamma_{MN}$.
Additionally, the coupling to the background form is dictated by the Wess--Zumino (WZ) action    
\be
S_{WZ} = -s\,\frac{T_{p+4}}{2}\int {\rm Str}\left(C_{p+1}\wedge F\wedge F\right) \ .
\ee
Here, $s^2=1$ is another sign to differentiate the coupling of branes and anti-branes. The induced metric on the branes, $g_{MN}$, and the dilaton, $\Phi$, depend on the scalar $Z$ describing the embedding through the warp factor 
\be
{\cal P}[H] = L^{\frac{7-p}{2}}\left( \delta_{ij}\, y^i  y^j +   Z\cdot Z \right)^{\frac{p-7}{2}}\ ,
\ee
and from now on we denote this pullback simply as $H$. The symbol $D$ represents a gauge-covariant derivative so that both $DZ$ and $F$ are U($\nf$) valued. Crucially, the symmetrised trace allows to treat them, as well as $Z$ appearing in the functional dependence of the background, as commuting under it \cite{Myers:1999ps}.\footnote{On the contrary, $Z$ and $A$ themselves, as well as their fluctuations, are considered \it{non-commuting}.}  Notice that some of the matrix identities used in the following are verified only for commuting coefficients, i.e., are valid only under the symmetrised trace.

To get the equations of motion we have to vary the total action with respect to the gauge fields and scalar. The variation of the DBI part is
\be\label{deltaDBI}
\delta S_{DBI} = -T_{p+4}\int\d^{p+1}x\d^4y \,{\rm Str}\left[\sqrt{-\det\Gamma}\left(\delta\left( e^{-\Phi}\right) + \frac{1}{2} e^{-\Phi} \left( \Gamma^{-1} \right)^{MN} \delta\Gamma_{NM} \right) \right] \ .
\ee

In order to compute this we need to invert the matrix $\Gamma$. Notice that once we have performed the variation we are allowed to use the BPS equations to prove that it vanishes, which facilitates the inversion of the matrix. The configuration we want to study is characterised by
\be\label{configuration}
F_{\mu\nu} = 0 \ , \qquad F_{\mu i} = E_i\left(y\right) \ \delta^t_\mu \ , \qquad F_{ij} = F_{ij}\left(y\right) \ , \qquad Z = Z\left(y\right)\ ,
\ee
so that none of the fields depends on the coordinates $\left\{t, \vec x\right\}$, and from the gauge components along those directions we are exciting just the temporal one $A_t(y)$. Moreover, for the solution to preserve supersymmetry we impose
\be\label{selfdual}
F_{ij} = \frac{\sigma}{2}\epsilon_{ijkl}F^{kl}\ , \qquad \sigma^2 = 1 \ ,
\ee
which is (anti)self-duality in the four-dimensional space along the flavor branes but transverse to the color ones. Notice that this equation is conformally invariant, so it is equivalent to impose (anti)self-duality in the flat space with flat metric $\delta_{ij }\d y^i \d y^j$. This condition has to be supplemented with 
\be\label{identification}
A_t = \tau Z \ , \qquad \tau^2 = 1 \qquad \Rightarrow \qquad D_\mu Z = 0 \ ,\qquad D_i Z = \tau F_{it} = - \tau E_i \ .
\ee
With this ansatz the inversion of $\Gamma$ gives the non-vanishing components
\be
\bal
\left( \Gamma^{-1} \right)^{x^m x^n} & = g^{x^m x^n} \ , \qquad  (m,n  =  1,\cdots,p) \\
\left( \Gamma^{-1} \right)^{tt} & = g^{tt}-H E_i E_j \left( \left[ g+F \right]^{-1} \right)^{ij} \ , \qquad \, \left( \Gamma^{-1}\right)^{ij} = \left( \left[ g+F \right]^{-1}\right)^{ij} \ , \\
\left( \Gamma^{-1} \right)^{it} & = H^{\frac{1}{2}} \left( \left[ g+F \right]^{-1} \right)^{ij} E_{j} \ , \qquad\qquad\quad \left( \Gamma^{-1} \right)^{tj} = -H^{\frac{1}{2}} E_i \left( \left[ g+F \right]^{-1} \right)^{ij} \ ,
\eal
\ee
To invert the $4\times4$ matrix $\left(g_{ij}+F_{ij}\right)$, we use that for any antisymmetric matrix it is verified
\be
F^{i k_1} \epsilon_{j k_1 k_2 k_3} F^{k_2 k_3} = \frac{1}{4} \delta^i{_j} \epsilon_{k_1 k_2 k_3 k_4} F^{k_2 k_2} F^{k_3 k_4}\ .
\ee
In conjunction with this, (anti)self-duality implies
\be
F^{ik} F_{kj} = \frac{1}{4} \delta^i{_j} F_{kl} F^{lk} \ ,
\ee
that is, up to a normalisation $F$ is its own inverse. Thus, the matrix we are looking for reads:
\be\label{eq.gplusFinverse}
\left( g+F \right)^{-1} = \frac{1}{\left( 1 - \frac{1}{4} F^{ij} F_{ji} \right)}\left( g-F \right) \ .
\ee

Finally,  the determinant of the matrix $\Gamma$ greatly simplifies
\be\label{determinant}
\bal
-\det \Gamma & = -\det g_{\mu\nu}\,\det\left( g_{ij} + H^{\frac{1}{2}} D_{\left(i \right.}Z D_{\left.j\right)}Z +F_{ij} - F_{i\mu}g^{\mu\nu}F_{\nu j}\right) \\
& = H^{-\frac{p+1}{2}}\det\left( g_{ij} + F_{ij} \right) = H^{\frac{3-p}{2}} \left( 1 - \frac{1}{4}  F^{ij} F_{ji} \right)^2 \ ,
\eal
\ee
where the first step is just the determinant in blocks, the second step is a cancelation between the terms containing $D_i Z$ and those with $F_{it}$ due to the configuration \eqref{identification},  and the last one follows from \eqref{eq.gplusFinverse} noticing that $\det(g+F)=\det(g-F)$.

The WZ piece gives simply
\be\label{deltaWZ}
\bal
\delta S_{WZ} & = T_{p+4}\frac{q\,s}{8} \delta\left[ \int \d^{p+1}x \d^4y \, {\rm Str} \left( \epsilon_{ijkl} F^{ij} F^{kl} \right)\right] \\
& = T_{p+4} \frac{q\,s\,\sigma}{2}  \int \d^{p+1}x \d^4y \, {\rm Str} \left( H^{\frac{1}{2}} F^{ij} \delta \left( H^{-\frac{1}{2}} F_{ij} \right) \right)  \\
& = T_{p+4}\frac{q\,s\,\sigma}{4} \int \d^{p+1}x \d^4y \, {\rm tr} \left( H^{-1} F_{ij}F^{ji} \delta H + 2 F^{ij} \delta F_{ij}\right) \ ,
\eal
\ee
with the first equality  obtained by substituting the value of the background form, Eq.~(\ref{Cp1}), while in the second we have used (anti)self-duality of the field strength.

\textbf{Variation with respect to the scalar:} The scalar $Z$ appears in two kinds of terms, those with $DZ$ and those with $Z\cdot Z$, coming from the warp factor. The variation with respect to the warp factor has several contributions. First, from the variation of the dilaton (first term in Eqn.~\eqref{deltaDBI}), which using the value for the determinant obtained in \eqref{determinant} is
\be
\bal
& - T_{p+4} \int \d^{p+1}x \d^4y \, \delta \left( e^{-\Phi} \right) \, {\rm Str} \left[ \sqrt{ -\det\Gamma} \right] \\
& = -T_{p+4} \, \frac{p-3}{4} \, \int \d^{p+1}x \d^4y \, H^{-1} {\rm tr} \left( 1 - \frac{1}{4} F^{ij} F_{ji} \right) \delta H \ .
\eal
\ee
Second, from the variation of the metric contained in $\delta\Gamma_{MN}$ (second term in Eqn.~\eqref{deltaDBI}) which within our ansatz turns out to be
\be\label{eq.variationwrtmetric}
\bal
& - T_{p+4} \int \d^{p+1}x \d^4y \, e^{-\Phi} \, {\rm Str} \left[ \sqrt{-\det\Gamma} \frac{1}{2} \left( \Gamma^{-1} \right)^{MN}\delta g_{NM} \right] \\
& = T_{p+4} \int \d^{p+1}x \d^4y \, H^{-1} {\rm tr} \left( \frac{p+1}{4} \left( 1 - \frac{1}{4} F^{ij} F_{ji} \right) + \frac{1}{4} H^{\frac{1}{2}} E_i E^i - 1 \right) \delta H \ .
\eal
\ee
If we combine these two with the term coming from the variation of the WZ, Eqn.~\eqref{deltaWZ}, we see that fixing the arbitrary signs as
\be
q \, s \, \sigma = 1
\ee
makes the variation with respect to the warp factor vanish, apart from the term proportional to $E_i E^i$. This is however canceled by a term coming from the variation of the kinetic term of the scalar
\be
\bal
& - T_{p+4} \int \d^{p+1}x \d^4y \, {\rm Str} \left[ \sqrt{ -\det\Gamma } \frac{1}{2} \left( \Gamma^{-1} \right)^{\left( MN \right)} \left( \delta \left( H^{\frac{1}{2}} \right) D_{M} Z D_{N} Z + 2 H^{\frac{1}{2}} D_{M} Z D_{N} \delta Z \right) \right] \\
& = -T_{p+4} \int \d^{p+1}x \d^4y \, {\rm Str} \left[ \frac{1}{4} H^{-\frac{1}{2}} D^i Z D_i Z \delta H + \delta^{ij} D_i Z D_j\delta Z - \tau H F^{ij} D_i Z D_j Z \left( D_t \delta Z \right) \right] \ .
\eal
\ee
The first term cancels with the $E_i E^i$ term in Eqn.~\eqref{eq.variationwrtmetric} upon using the BPS condition (\ref{identification}), as promised. The last term vanishes because of symmetry, since as we mentioned $F$ and $DZ$ commute under the symmetrised trace. Thus, after integration by parts we are left with the equation for the scalar:
\be\label{eomscalar}
\delta^{ij} D_ i D_j Z = 0 \ ,
\ee
which is the Laplace equation {\it on the flat space} $\delta_{ij} \d y^i \d y^j$ but containing the gauge connection. 

\textbf{Variation with respect to the gauge field:} The kinetic term of the scalar also contains a variation of the gauge field, that in our ansatz reduces to
\be
\bal
& - T_{p+4} \int \d^{p+1}x \d^4y \, H^{\frac{1}{2}} \, {\rm Str} \left( \sqrt{-\det\Gamma} \left( \Gamma^{-1} \right)^{\left(MN\right)} D_{M}Z \left[ \delta A_{N},Z \right] \right) \\
& = - T_{p+4} \int \d^{p+1}x \d^4y \, H^{\frac{1}{2}} \, {\rm Str} \left( \delta A_i \left[Z,D^i Z\right] \right) \ ,
\eal
\ee
where we have discarded an additional piece for the same symmetry reasons as before. Finally, we have the contributions coming from the variation of the kinetic term of the gauge field, that read
\be
\bal
& T_{p+4} \int \d^{p+1}x \d^4y \, {\rm Str} \left[ \sqrt{-\det\Gamma} \left( \Gamma^{-1} \right)^{MN} D_{[M}\delta A_{N]}\right] \\
& = T_{p+4} \int \d^{p+1}x \d^4y \, {\rm Str} \left( \delta A_j D_i F^{ij} - \tau \delta A_t \delta^{ij}D_i D_j Z + \tau H^{\frac{1}{2}} \delta A_i D_t D^i Z \right) \ ,
\eal
\ee
where we have integrated by parts in the last step. Variation with respect to the temporal component gives again the equation of motion of the scalar (\ref{eomscalar}), while the magnetic components must solve
\be
D_i F^{ij} + \tau H^{\frac{1}{2}} D_t D^j Z - H^{\frac{1}{2}} \left[ Z, D^j Z \right] = D_i F^{ij} = 0 \ ,
\ee
where we have used that for an adjoint field $D_t=\partial_t+\left[A_t,\cdot\right]=\partial_t+\tau\left[Z,\cdot\right]$. This equation is trivially verified for (anti)self-dual configurations due to the Bianchi identity. For the same reason, the last term in the variation of the WZ action, Eq.~(\ref{deltaWZ}), also vanishes. 

In summary, the variation of the non-Abelian action of \cite{Myers:1999ps} vanishes for configurations of the form (\ref{configuration}) if the self-duality equation (\ref{selfdual}), together with (\ref{eomscalar}) for the scalar and the identification (\ref{identification}), are verified. Notice that we do not need to specify any direction in gauge space for $Z$ nor $A_t$ as long as they are collinear. The proof is valid both for self and anti self-dual configurations and arbitrary rank $\nf$.

\section{The SU(2) instanton as a lego for higher-rank gauge groups\label{app.su2}}

In this section, we give the solution to (\ref{selfdual}) and (\ref{eomscalar}) corresponding to the dyonic instanton of \cite{Lambert:1999ua}. The equations can be solved sequentially, since (anti)self-duality only involves the magnetic components of the gauge field, while $DZ$ requires knowledge of the gauge potential. We give the solution for arbitrary gauge group and this requires rewriting the associated algebra in canonical form.

Consider $\mathfrak{g}$, a semi-simple Lie algebra of finite dimension.\footnote{One could extend the results to simple algebras, but we do not need it for our purposes since the Abelian ideal, corresponding to the baryonic chemical potential, vanishes in our solutions.} A Cartan subalgebra $\mathfrak{h}\subset\mathfrak{g}$ is a maximal Abelian subalgebra such that all its elements are diagonalisable in the adjoint representation. Its dimension is the rank $r$ of $\mathfrak{g}$, with $r=n-1$ for $\mathfrak{su}\left(n\right)$. 

Let us choose a basis $H_{\hat \imath}$, ${\hat \imath}=1,\dots,r$ of $\mathfrak{h}$, and let these generators be Hermitian for convenience. Since by definition they form an Abelian subalgebra we have
\be
\left[H_{\hat \imath},H_{\hat \jmath}\right] = 0 \ .
\ee
This can be completed to a basis for the algebra by finding the eigenvectors of ${\rm ad}H_{\hat \imath}=\left[H_{\hat \imath},\cdot\right]$ that are linearly independent from the $H_{\hat \imath}$, that is, a set of elements $E_\alpha$ of $\mathfrak{g}$ verifying 
\be
\left[H_{\hat \imath},E_\alpha\right] = \alpha_{\hat \imath} E_\alpha \ .
\ee
Notice that not all $\alpha_{\hat \imath}$ can vanish simultaneously, since we assume $E_\alpha\notin\mathfrak{h}$. The real eigenvalues $\alpha=\left(\alpha_1,\dots,\alpha_r\right)$ are called roots, have no multiplicity and there are $d-r$ of them in a given $d$-dimensional Lie algebra. Moreover, if $\alpha$ is a root then $-\alpha$ is also a root with generator $E_{-\alpha}=E_\alpha^\dagger$, so roots always come in pairs. Indeed, it can be shown that
\be
\left[E_\alpha,E_{-\alpha}\right] = \alpha_{\hat \imath} H_{\hat \imath}\equiv H_\alpha \ , 
\ee
where we have normalised so that the non-vanishing components of the Killing form are
\be
{\rm tr}\left(H_{\hat \imath} H_{\hat \jmath}\right) = \delta_{{\hat \imath}{\hat \jmath}} \ , \qquad {\rm tr}\left(E_\alpha E_{-\alpha}\right) = 1 \ .
\ee
Finally, the commutator between generators associated to different roots is
\begin{equation}\label{EaEb}
\left[E_\alpha,E_\beta\right] = N_{\alpha,\beta}E_{\alpha+\beta}\,
\end{equation}
where $N_{\alpha,\beta}$ is a constant that vanishes if $\alpha+\beta$ is not a root. We will give the explicit expression and properties of these constants when needed. 
 
In the simplest case of $\mathfrak{su}\left(2\right)$, whose rank is $r=1$, the only roots are $\alpha=\sqrt{2}$ together with its opposite. The generators $E_{\pm\alpha}$ are the usual raising and lowering operators, familiar from the treatment of spin. These, together with the unique Cartan, are constructed from the Pauli matrices as 
\begin{equation}
E_{\pm\alpha} = \frac{1}{2}\left(\sigma^1\pm i\, \sigma^2\right) \ , \qquad\qquad H = \frac{1}{\sqrt{2}}\sigma^3 \ .
\end{equation}
For larger algebras, there is an $\mathfrak{su}\left(2\right)$ subalgebra $\left\{E_{\pm\alpha},H_\alpha\right\}$ associated to each non-zero pair of root vectors $\pm\alpha$. Thus, we can use any of these subalgebras to construct the basic SU(2) instanton. 

In the following we consider only the solutions obtained from 't Hooft's ansatz, adapted to this language. Let us define the antisymmetric tensors
\begin{equation}
\overline{\eta}^{\pm}_{ij} = \frac{1}{\sqrt{2}}\left(\overline{\eta}^1_{ij}\pm i\,\overline{\eta}^2_{ij}\right) \ , \qquad\qquad \overline{\eta}^H_{ij} = \overline{\eta}^3_{ij} \ ,
\end{equation} 
constructed from the usual anti self-dual 't Hooft tensors $\overline{\eta}^a$, $a=1,2,3$. Inherited from their properties, we have that
\be\label{etaproperties}
\bal
\overline{\eta}^H_{ik}\overline{\eta}^H_{jk} & = \delta_{ij} \ ,  \hspace{15.4em} \overline{\eta}^\pm_{ik}\overline{\eta}^\pm_{jk} = 0 \ , \\
\overline{\eta}^\pm_{ik}\overline{\eta}^H_{jk} & = \pm i\,\overline{\eta}^\pm_{ij} \ , \hspace{14em}  \overline{\eta}^\pm_{ik}\overline{\eta}^\mp_{jk} = \delta_{ij} \mp i\,\overline{\eta}^H_{ij} \ , \\
\overline{\eta}^\pm_{ij}\overline{\eta}^H_{kl} - \overline{\eta}^H_{ij}\overline{\eta}^\pm_{kl} & = \pm i \left(\delta_{ik}\overline{\eta}^\pm_{jl}+\delta_{jl}\overline{\eta}^\pm_{ik}-\delta_{il}\overline{\eta}^\pm_{jk}-\delta_{jk}\overline{\eta}^\pm_{il}\right) \ , \\
\overline{\eta}^+_{ij}\overline{\eta}^-_{kl}-\overline{\eta}^-_{ij}\overline{\eta}^+_{kl} & = - i \left(\delta_{ik}\overline{\eta}^H_{jl}+\delta_{jl}\overline{\eta}^H_{ik}-\delta_{il}\overline{\eta}^H_{jk}-\delta_{jk}\overline{\eta}^H_{il}\right) \ , \\
\epsilon_{ijk}{}^s\overline{\eta}^{\pm,H}_{sl} & = \delta_{il}\overline{\eta}^{\pm,H}_{jk}-\delta_{jl}\overline{\eta}^{\pm,H}_{ik}+\delta_{kl}\overline{\eta}^{\pm,H}_{ij} \ .
\eal
\ee
Using these tensors, we take the following ansatz for the gauge potential
\be
A_i = \left( \frac{1}{|\alpha|} \overline{\eta}^+_{ij} \otimes E_{-\alpha} + \frac{1}{|\alpha|} \overline{\eta}^-_{ij} \otimes E_{+\alpha} + \frac{1}{|\alpha|^2} \overline{\eta}^H_{ij} \otimes H_\alpha\right) \partial_j \log \varphi \ , 
\ee
with $|\alpha|^2 = \alpha\cdot\alpha = \alpha_{\hat \imath}\alpha_{\hat \imath}$.
Then, using the properties for the $\overline{\eta}$ listed in (\ref{etaproperties}), it can be seen that the field strength inherited from this gauge potential is self-dual whenever $\varphi$ solves 
\be
\frac{\delta^{ij}\partial_i \partial_j \varphi}{\varphi} = 0 \ .
\ee
A similar set of self-dual tensors provides an anti self-dual field strength given the same condition on $\varphi$. 

Once we have an (anti)self-dual field strength, we need to solve (\ref{eomscalar}) in its background for the embedding. Remember that $Z$ can be pointing in any direction in gauge space. There are two distinct possibilities (and linear combinations thereof): we can place the embedding in the $\mathfrak{su}(2)$ subalgebra of the instanton or in a different one, corresponding to a different root. In the first case we can use $\mathfrak{su}(2)$ rotations to align the embedding in the direction of the Cartan, so we take the ansatz 
\be\label{Zalpha}
Z = f\left(\varphi\right)\,H_\alpha \ .
\ee
Again, exploiting the properties of the $\overline{\eta}$ tensors it can be shown that the equation for the scalar is solved if 
\be
\varphi^2 f'' = 2f\qquad\Rightarrow \qquad f\left(\varphi\right) = \frac{M_1}{\varphi}+M_2\,\varphi^2 \ .
\ee
with $'$ denoting derivative with respect to $\varphi$ and $M_1$ and $M_2$ a pair of (real) integration constants. On the other hand, we could take the embedding along the raising and lowering operators  corresponding to a different root, that is,
\be
Z = f_+\left( \varphi \right) \, E_{+\beta} + f_- \left( \varphi \right) E_{-\beta} \ .
\ee
In our conventions the embedding has to be hermitean, forcing $f_+^*=f_-$. Similar manipulations imply that the equation for the $E_{+\beta}$ component is solved as long as
\be\label{Eqf+}
\varphi^2 f_+'' = \frac{f_+}{|\alpha|^2}\left(N_{\alpha,\beta}N_{-\alpha,\alpha+\beta}+N_{-\alpha,\beta}N_{\alpha,-\alpha+\beta}+\frac{\left(\alpha\cdot\beta\right)^2}{|\alpha|^2}\right) \ .
\ee
Crucially, this equation can be solved for any Lie algebra, given any pair of roots $\alpha$ and $\beta$. It can be easily proven that the real coefficients $N_{\alpha,\beta}$ verify the following relations
\be
N_{\alpha\beta} = -N_{\beta,\alpha} = -N_{-\alpha,-\beta} \ , 
\ee
and, from the Jacobi identity applied to the generators associated to three roots satisfying $\alpha+\beta+\gamma=0$, that
\be
N_{\alpha,\beta} = N_{\beta,-\alpha-\beta} = N_{-\alpha-\beta,\alpha} \ .
\ee
As a consequence of these 
\be
N_{-\alpha,\alpha+\beta} = N_{\alpha,\beta} \ .
\ee
On the other hand, the Jacobi identity applied to $E_{+\alpha}$, $E_{-\alpha}$ and $E_{+\beta}$ forces the relation
\be
\alpha\cdot\beta = N_{-\alpha,\beta}N_{\alpha,\beta-\alpha}+N_{\beta,\alpha}N_{-\alpha,\alpha+\beta} = N_{-\alpha,\beta}^2-N_{\alpha,\beta}^2 \ .
\ee
Now, let $p\le0$ be the smallest integer such that $\alpha+p\beta$ is a root and let $q\ge0$ be the largest integer such that $\alpha+q\beta$ is a root. One of the most basic properties of Lie algebras is that these integers verify
\be
2\,\frac{\alpha\cdot\beta}{|\alpha|^2} = -\left(p+q\right) \ .
\ee
Moreover, the coefficients in the commutator (\ref{EaEb}) are given in terms of them as 
\be
N_{\alpha,\beta}^2 = \frac{1}{2}\left(1-p\right)q|\alpha|^2 \ , 
\ee
which vanishes only if $q=0$, that is, if $\alpha+\beta$ is not a root. Using all these results, (\ref{Eqf+}) is solved by
\be
f_+ = M_1\,\varphi^{\frac{p-q}{2}} + M_2\,\varphi^{\frac{q-p+2}{2}} \ .
\ee
Notice that this is invariant under $q\to-p$ and $p\to-q$ simultaneously, which corresponds to changing $\beta$ into $-\beta$. This means that $f_-$ has a solution of the same form, while hermiticity selects the integration constants so that 
\be
f_- = M_1^*\,\varphi^{\frac{p-q}{2}}+M_2^*\,\varphi^{\frac{q-p+2}{2}} \ .
\ee

Finally we can align the embedding with the Cartan associated to a different root $\beta$. However it is easy to see that this produces a vanishing solution unless $\alpha\cdot\beta=0$. More generally, we can combine this embedding with (\ref{Zalpha}), that is, we can take
\be
Z = f_\alpha \left( \varphi\right) \, H_\alpha + f_\beta\left( \varphi \right) H_\beta \ .
\ee 
Then, the functions have to satisfy 
\be
\begin{array}{rclcrcl}
 f_\beta''&=& 0 &\qquad\Rightarrow&\qquad f_\beta&=&M_1+M_2\,\varphi \ , \\[2mm]
 \varphi^2f_\alpha''&=&2\left(f_\alpha+\frac{\alpha\cdot\beta}{|\alpha|^2}f_\beta\right)&\qquad\Rightarrow&\qquad f_\alpha&=&\frac{M_3}{\varphi}+M_4\,\varphi^2+\frac{p+q}{2}f_\beta \ , 
\end{array}
\ee
where we see that $f_\alpha=0$ is not consistent with non-trivial $f_\beta$ unless $\alpha\cdot\beta=0$, as mentioned above.

We have thus reduced the problem to compute the possible values for $p$ and $q$ for a given Lie algebra. Coming back to the system of branes, we know that the gauge symmetry associated to the flavor branes is U($\nf$). Since we are not exciting the ${\rm U}(1)_\mt{B}\subset {\rm U}(\nf)$, we can apply the results above to the algebra $\mathfrak{su}\left(\nf\right)$, whose roots system is well known. Indeed, for any rank, all the roots have the same length, $|\alpha|^2=2$, and the only possible non-trivial values for the integers are
\be
p=0 \ , \quad\quad q=1\,\quad\quad{\rm or}\quad\quad p=-1 \ , \quad\quad q=0 \ .
\ee
These give the same embedding, since they are related by the symmetry discussed above. 

The single instanton solution corresponds to the function
\be\label{1instanton}
\varphi = \frac{r^2+\Lambda^2}{r^2} \ , \qquad r^2 = \sum_{i=1}^4y_i^2 \ , 
\ee
so the possible regular embeddings are
\be
Z_\alpha = \Mq \frac{r^2}{r^2+\Lambda^2} H_\alpha \ , \qquad Z_\beta = \frac{r}{\sqrt{r^2+\Lambda^2}}\left(\Mq \, E_{+\beta} + \Mq^* \, E_{-\beta} \right) \ .
\ee

\section{Fluctuation channels\label{app.fluctuations}}

In this appendix we provide the equations of motion for the fluctuations present in our model, classified according to their transformation under the global symmetry group SO(2)$\times$U(1)$_\mt{D}$ and local symmetry properties. Later we will give more details about the behavior of the fields and the numerical procedure we implemented for the two channels we have focused on in the main text.

\begin{enumerate}

\item \textbf{SO(2) vectors, charged and uncharged under U(1)$_\iso$}: These are two channels consisting of two exact copies of the same fluctuation given by the fields $\alpha_{y \hat a}$ and $\alpha_{z \hat a}$. Focusing on the former we can write the following combinations charged under U(1)$_\iso$
\be
\alpha_{y \pm} = \alpha_{y1} \pm i \alpha_{y2} \ ,
\ee
with equations of motion
\be
\alpha_{y\pm}'' + \frac{3}{r} \alpha_{y\pm} ' - \left( \frac{8\, a^2}{r^2} + \frac{L^4 \left( k^2 - \left( \omega \pm 2 a_t \right)^2 + 4 \phi^2\right)}{\left( r^2 + \phi^2 \right)^2 } \right) \alpha_{y\pm} = 0 \ .
\ee

The fluctuation $\alpha_{y3}$ is neutral under the U(1) factors and its equation of motion reads
\be
\alpha_{y3}'' + \frac{3}{r} \alpha_{y3} ' - \left( \frac{8\, a^2}{r^2} + \frac{L^4 \left( k^2 - \omega^2 \right)}{\left( r^2 + \phi^2 \right)^2 } \right) \alpha_{y3} = 0 \ .
\ee

\item \textbf{Neutral SO(2) scalar}: This channel comprises only the $\beta_3$ fluctuation, which is a scalar under SO(2) and a singlet under both U(1) groups.
The equation of motion is
\be\label{eq.beta3eom}
\beta_3'' + \frac{3}{r} \beta_3' - \left( \frac{8 \, a^2}{r^2} + \frac{L^4 \left( k^2 - \omega^2 \right) }{ \left( r^2 + \phi^2 \right)^2 } \right) \beta_3 = 0 \ .
\ee

\item \textbf{SO(2) scalars charged under U(1)$_\iso$ and independent of gauge transformations}: In this channel we encounter the fluctuations $\beta_1$ and $\beta_2$, which are rotated onto each other by action of the group U(1)$_\iso$. The fluctuations are better expressed with the combinations
\be
\beta_\pm = \beta_1 \pm i\, \beta_2
\ee
in terms of which
\be\label{eq.beta12eoms}
\beta_\pm'' + \frac{3}{r} \beta_\pm' - \left( \frac{8 \, a^2}{r^2} + \frac{L^4 \left( k^2 - \left( \omega \pm 2 a_t \right)^2 + 4 \phi^2 \right) }{ \left( r^2 + \phi^2 \right)^2 } \right) \beta_\pm = 0
\ee

\item \textbf{SO(2) scalars neutral under the diagonal of U(1)$_\Rsym \times$U(1)$_\iso$ and even}: This channel contains the fluctuations that do not change under action of the diagonal of U(1)$_\Rsym \times$U(1)$_\iso$ and carry none or one index of each one. The gauge transformation given by $\lambda_3$ in \eqref{eq.gaugetransformation} acts in this channel as
\be\label{eq.gaugelambda3}
\alpha_{t3} = - i \omega \lambda_3 \ ,  \quad \alpha_{x3} =  i k \lambda_3 \ , \quad \alpha_{12} - \alpha_{21} = 4 \, a(r) \lambda_3 \ .
\ee
Obvious perturbations belonging to this channel are then $\alpha_{t3}$, $\alpha_{x3}$, and $\alpha_{12} - \alpha_{21}$. The latter combination is not neutral under U(1)$_\Rsym$ or U(1)$_\iso$ individually, but only to the simultaneous acting of both. A second combination of $\alpha_{a \hat a}$ fluctuations that is also invariant under the diagonal of U(1)$_\Rsym \times$U(1)$_\iso$ is $\alpha_{11}+\alpha_{22}$, and in fact it is convenient to define
\be\label{eq.alphapm}
\alpha_\pm = \alpha_{11} + \alpha_{22} \pm i \left( \alpha_{12} - \alpha_{21} \right) \ .
\ee
It is not hard to convince oneself that, since $\alpha_{11}+\alpha_{22}$ and $\alpha_{t3}$ are fluctuations of fields present in the background, this channel also captures the fluctuations $\alpha_{33}$ and $\zeta_3$, which are neutral under  U(1)$_\Rsym$ and U(1)$_\iso$ separately and, therefore, also under its diagonal combination. Finally, the radial component $\alpha_{r3}$ also belongs to this channel, and keeping it to zero provides a first-order constraint.

\be
\bal
& \alpha_{t3}'' + \frac{3}{r} \alpha_{t3}' - L^4 \frac{k (k \alpha_{t3}+\omega \alpha_{x3})}{(r^2 + \phi^2)^2} - \frac{8a^2}{r^2} \alpha_{t3} - \frac{a}{r^2} \left( 4 a_t (\alpha_++\alpha_-) + \omega (\alpha_+ - \alpha_-) \right) = 0 \ , \\
& \alpha_{x3}'' + \frac{3}{r} \alpha_{x3}' + L^4  \frac{\omega (k \alpha_{t3}+\omega \alpha_{x3})}{(r^2 + \phi^2)^2} - \frac{8a^2}{r^2} \alpha_{x3} + \frac{a}{r^2}  k (\alpha_+ - \alpha_-) = 0\ , \\
& \zeta_3'' + \frac{3}{r} \zeta_3' - L^4  \frac{(k^2-\omega^2)}{(r^2+\phi^2)^2} \zeta_3 - \frac{8a^2}{r^2}\zeta_3 - \frac{4 a \phi}{r^2} (\alpha_+ + \alpha_-) = 0 \ , \\
& \alpha_{33}'' + \log ' \left[ r(r^2+\phi^2)^2) \right] \alpha_{33}' - \left( L^4 \frac{(k^2-\omega^2)}{(r^2+\phi^2)^2} + \frac{8a^2}{r^2} - \frac{4}{r^2} \frac{r^2 - \phi^2 + 2 r \phi \phi'}{r^2+\phi^2}\right) \alpha_{33} \\
& \quad - \frac{2a}{r^2} \frac{2a(r^2+\phi^2)-2r\phi\phi'+ r^2 + 3\phi^2}{r^2+\phi^2}(\alpha_+ +\alpha_-) = 0 \ , \\
& \alpha_\pm''  +  \log ' \left[ r(r^2+\phi^2)^2) \right] \alpha_\pm'- L^4  \frac{ k^2- \left( \omega \pm 2 a_t \right)^2 + 4\phi^2 }{(r^2+\phi^2)^2} \alpha_{\pm} - \frac{12a^2}{r^2} \alpha_{\pm} - \frac{4a^2}{r^2} \alpha_\mp  \\
& \quad + \frac{4}{r^2} \frac{r^2 - \phi^2 + 2 r (1+a)\phi \phi'}{r^2+\phi^2} \alpha_{\pm} + \frac{4 L^4 (a_t^2 - \phi^2)}{(r^2 + \phi^2)^2} \mp \frac{16 L^4 a\phi}{(r^2+\phi^2)^2} \zeta_3 \\
& \quad  + \frac{4 L^4 a}{(r^2 + \phi^2)^2} \left( k \alpha_{x3} + \left(\omega \pm 4 a_t \right) \alpha_{t3} \right) - \frac{8a}{r^2} \frac{2a(r^2+\phi^2) - 2r\phi\phi' + r^2 + 3\phi^2}{r^2+\phi^2} \alpha_{33} = 0 \ , 
\eal
\ee
and the constraint
\be
 L^4 r^3(\omega \alpha_{t3}'+k \alpha_{x3}') + 2 a (1+a) (r^2+\phi^2)^2 (\alpha_+ - \alpha_-) + a \,r(r^2+\phi^2)^2 (\alpha_+' - \alpha_-') = 0 \ .
\ee

\item \textbf{SO(2) scalars charged under the diagonal of U(1)$_\Rsym \times$U(1)$_\iso$ and even}: These two fluctuations carry two indices, one for each U(1) group, and under the diagonal product they are charged. As usual, they are better expressed as the combinations
\be\label{eq.alphapmhat}
\widehat \alpha_\pm = \alpha_{11} - \alpha_{22} \pm i \left( \alpha_{12} + \alpha_{21} \right) \ ,
\ee
in terms of which the equations of motion read
\be
\bal
& \widehat \alpha_\pm''  + \partial_r \log \left[ r(r^2+\phi^2)^2) \right] \widehat \alpha_\pm'- L^4 \frac{k^2 - \left( \omega \pm 2 a_t \right)^2 + 4 \phi^2}{(r^2+\phi^2)^2} \widehat \alpha_{\pm} + \frac{4}{r^2} \frac{r^2 - \phi^2 + 2 r \phi \phi'}{r^2+\phi^2} \widehat \alpha_{\pm} \\
& \quad - \frac{4a}{r^2} \frac{2r\phi\phi'-r^2-3\phi^2}{r^2+\phi^2} \widehat \alpha_\pm = 0 \ , 
\eal
\ee

\item \textbf{SO(2) scalars neutral under the diagonal of U(1)$_\Rsym \times$U(1)$_\iso$ and odd}: The fluctuations encountered in this channel are $\alpha_{t1}$, $\alpha_{t2}$, $\alpha_{x1}$, $\alpha_{x2}$, $\alpha_{13}$, $\alpha_{23}$, $\alpha_{31}$, $\alpha_{32}$, $\zeta_{1}$ and $\zeta_2$. The radial components $\alpha_{r1}$ and $\alpha_{r2}$ also belong here, and keeping them to zero provide two first-order constraints. Notice that gauge transformation given by $\lambda_1$ and $\lambda_2$ in \eqref{eq.gaugetransformation} act in this channel as
\be\bal\label{eq.gaugelambda12}
& \alpha_{t1} = - i \omega \lambda_1 + 2\, a_t(r) \lambda_2 \ ,  \quad \alpha_{x1} =  i k \lambda_1 \ , \quad \alpha_{23} - \alpha_{32} =  4\, a(r) \lambda_1 \ , \quad \zeta_2=- 2\,\phi(r)\, \lambda_1 \ , \\
& \alpha_{t2} = - i \omega \lambda_2 - 2\, a_t(r) \lambda_1 \ ,  \quad \alpha_{x2} =  i k \lambda_2 \ , \quad \alpha_{13} - \alpha_{31} =  -4\, a(r) \lambda_2 \ , \quad \zeta_1= 2\, \phi(r)\, \lambda_2  \ .
\eal\ee

The set of second-order equations of motion reads
\be
\bal
& \alpha_{tI}'' + \frac {3} {\rho} \alpha_{tI}' - \frac{4 L^4 \phi^2}{(r^2 + \phi^2)^2} \alpha_{tI} - L^4  \frac{k (k \alpha_{tI}+\omega \alpha_{xI})}{(r^2 + \phi^2)^2} - \frac{8a^2}{r^2} \alpha_{tI} -  \frac{2 L^4 ik a_t}{(r^2+\phi^2)^2} \varepsilon_I{^J} \alpha_{xJ}  \\
& \quad +  \frac{4 L^4 \phi a_t}{(r^2+\phi^2)^2} \zeta_I + \frac{2i\omega a_t}{(r^2+\phi^2)^2} \varepsilon_I{^J}\zeta_J +  \frac{4  L^4 a_t a}{r^2}  (\alpha_{I3} + \alpha_{3I}) \\
& \quad - \frac{2i\omega a}{r^2} \varepsilon_I{^J} (\alpha_{J3} - \alpha_{3J}) = 0\ , \\
& \alpha_{xI}'' + \frac{3}{r} \alpha_{xI}' + L^4  \frac{\omega (k \alpha_{tI}+\omega \alpha_{xI})}{(r^2 + \phi^2)^2} - \frac{8a^2}{r^2} \alpha_{xI} +  \frac{4 L^4 i\omega a_t}{(r^2+\phi^2)^2} \varepsilon_I{^J} \alpha_{xJ} \\
& \quad  +   \frac{4  L^4 \left(a_t^2 - \phi^2\right)}{( r^2 + \phi^2 )^2} \alpha_{tI}  +   \frac{2 L^4 ik a_t}{(r^2+\phi^2)^2} \varepsilon_I{^J} \alpha_{tJ} -   \frac{2 L^4 ik a_t}{(r^2+\phi^2)^2} \varepsilon_I{^J}\zeta_J \\
& \quad + \frac{2ik a}{r^2} \varepsilon_I{^J} (\alpha_{J3} - \alpha_{3J})  = 0 \ , \\
& \zeta_I'' + \frac {3} {\rho} \zeta_I - \frac{8a^2}{r^2} \zeta_I -  L^4 \frac{k^2 - \omega^2 - 4  a_t^2}{(r^2 + \phi^2)} \zeta_I + \frac{4 L^4 i\omega a_t}{(r^2+\phi^2)^2} \varepsilon_I{^J} \zeta_J + \frac{4a\phi}{r^2} \left( \alpha_{I3} + \alpha_{3I} \right)  \\
& \quad -  \frac{4 L^4 \phi a_t}{(r^2+\phi^2)^2} \alpha_{tI} - \frac{2 L^4 i\phi}{(r^2+\phi^2)^2} \varepsilon_I{^J} \left( \omega \alpha_{tJ} + k \alpha_{xJ} \right) = 0 \ , \\
& \alpha_{I3}'' + \log' \left[ r(r^2+\phi^2)^2) \right] \alpha_{I3}'-\left(  L^4 \frac{(k^2-\omega^2)}{(r^2+\phi^2)^2} + \frac{4a^2}{r^2} - \frac{4}{r^2} \frac{r^2 - \phi^2 + 2 r \phi \phi'}{r^2+\phi^2} \right) \alpha_{I3}  \\
& \quad + \frac{4a}{r^2} \frac{a(r^2+\phi^2)-2r\phi\phi'+r^2+3\phi^2}{r^2+\phi^2} \alpha_{3I} +  \frac{4 L^4  a}{(r^2+\phi^2)^2} \left(\phi \zeta_I - a_t \alpha_{tI} \right) \\
& \quad + \frac{2i L^4 a}{(r^2 + \phi^2)^2} \varepsilon_I{^J}\left( \omega \alpha_{tJ} + k \alpha_{xJ} \right) = 0 \ , \\
& \alpha_{3I}'' + \log' \left[ r(r^2+\phi^2)^2) \right] \alpha_{3I}'- \left(  L^4 \frac{(k^2-\omega^2)}{(r^2+\phi^2)^2} + \frac{4a^2}{r^2}- \frac{4}{r^2} \frac{r^2 - \phi^2 + 2 r \phi \phi'}{r^2+\phi^2}\right) \alpha_{3I} \\ 
& \quad + \frac{4(a_t^2 - \phi^2)}{(r^2+\phi^2)^2} \alpha_{3I}  + \frac{4a}{r^2} \frac{a(r^2+\phi^2)-2r\phi\phi' + r^2 + 3\phi^2}{r^2+\phi^2} \alpha_{I3} + \frac{4 L^4 i \omega  a_t}{(r^2+\phi^2)^2} \varepsilon_I{^J} \alpha_{3J} \\
& \quad + \frac{4 L^4  a}{(r^2+\phi^2)^2} \left(\phi \zeta_I - a_t \alpha_{tI} \right) - \frac{2 L^4 ia}{(r^2 + \phi^2)^2} \varepsilon_I{^J}\left( \omega \alpha_{tJ} + k \alpha_{xJ} \right) = 0
\eal
\ee
where $I,J=1,2$ and the antisymmetric matrix has entries $\varepsilon_1{^2}=1$ and $\varepsilon_2{^1}=-1$. The constraints are
\be
\bal
&  L^4 \left(r^3  a_t (\alpha_{tI}' - \zeta_I') - r^3  ( a_t' \alpha_{tI} - \phi' \zeta_I) + i\frac{r^3}{2} \varepsilon_I{^J} \left( \omega \alpha_{tJ}'+k \alpha_{xJ}' \right) \right) \\
& \quad + 2 a (1 + a) (r^2+\phi^2)^2 (\alpha_{I3} - \alpha_{3I}) + a \,r(r^2+\phi^2)^2 (\alpha_{I3}' - \alpha_{rI}') = 0 \ .
\eal
\ee

\end{enumerate}

\subsection{Neutral SO(2) scalar\label{app.beta3fluctuation}}

This channel is given by a single, decoupling fluctuation $\beta_3$, with equation of motion given in Eqn.~\eqref{eq.beta3eom}. In the instantonless limit, $a=0$ and $\phi=M_q$, the equation reduces to the one in \cite{Kruczenski:2003be}, which implies a discrete spectrum of regular, normalisable solutions
\be
\omega^2 - k^2 = \frac{\Mq^2}{L^4} 4 (n+1)(n+2)  \qquad (\text{with }n\geq0) \ .
\ee

For simplicity we write the remaining expressions in this subsection in terms of the dimensionless variables
\be\label{eq.rescaling}
(\omega ,k)= \frac{\Mq}{L^2}(\overline \omega, \overline k) \ , \qquad r = \Mq \overline r \ , \qquad \Lambda =  \Mq \, \overline \Lambda \ .
\ee
From the equation of motion we obtain the UV and IR asymptotics
\be\bal
\overline \beta_3 & = \overline  s + \frac{\overline  v}{\overline  r^2} + \frac{1}{2} (\overline \omega^2 - \overline k^2) \overline s \frac{\log[\overline r]}{\overline r^2} - \frac{3}{64} (\overline \omega^2 - \overline k^2)^2  \frac{\overline s}{\overline r^4} - \frac{(\overline \omega^2 - \overline k^2)^2}{16} \, \overline s \, \frac{\log[\overline r]}{\overline r^4} + {\mathcal{O}}(\overline r^{-6}\log[\overline r]) \\
\overline \beta_3 & = e^{- \frac{\sqrt{\overline k^2 - \overline \omega^2}}{\overline r}} \frac{\beta_3^{(0)}}{\overline r^{1/2}} \left( 1 + i \frac{35\overline \Lambda^4 + 8 (\overline \omega^2 - \overline k^2)}{8 \,  \overline \Lambda^4 \sqrt{\overline \omega^2 - \overline k^2}} \overline r + {\mathcal{O}}(\overline r^2) \right) \ ,
\eal\ee
respectively, and where $\beta_3 = \Mq \overline \beta_3$. We have already imposed an ingoing wave at the Poincar\'e horizon (see for example \cite{Son:2002sd}).

The spectral function is built by shooting from the IR with an arbitrary normalisation $\beta_3^{(0)}$, and reading the UV parameters $\overline v=\overline v(\overline \Lambda, \overline \omega^2 - \overline k^2)$ and $\overline s=\overline s(\overline \Lambda, \overline \omega^2 - \overline k^2)$. The spectral function is given, up to a constant, by
\be
\chi \propto \mathrm{Im}\, \frac{\overline v(\overline \Lambda, \overline \omega^2 - \overline k^2)}{\overline s(\overline \Lambda,\overline \omega^2 - \overline k^2)} \ .
\ee

\subsection{SO(2) scalars neutral under the diagonal of U(1)$_\Rsym \times$U(1)$_\iso$ and even\label{app.gaugelambda3}}

This section describes how to calculate the quasi-normal modes in the channel coupling the following fluctuations: $\alpha_{t3}$, $\alpha_{x3}$, $\alpha_+$, $\alpha_-$, $\alpha_{33}$ and $\zeta_3$, subjected to a first-order constraint. There are 6 fluctuations coupled in this channel, and each of these has an associated second order differential equation and there is one first-order constraint arising from the holographic gauge, so in total one expects 11 independent solutions. Of these, five are ingoing waves at the Poincar\'e horizon and five to be outgoing waves. The eleventh solution is a pure gauge solution to the equations of motion given by \eqref{eq.gaugelambda3}. This arises since we are not working with gauge invariant combinations, which introduces a redundancy that we have to deal with as indicated below.

The UV asymptotics of the fields are
\be\bal\label{eq.UVcoupled1}
\zeta_{3}&=s_3 + \frac{v_3}{r^2} + \cdots \ , \\
\alpha_{I}&=s_I + \frac{v_I}{r^2} + \cdots  \qquad \qquad\quad (\text{with }I=\{t3,x3\}) \ , \\
\alpha_{J}&= \frac{s_I}{r^2} \log[r] + \frac{v_J}{r^2} + \cdots  \qquad\, (\text{with }J=\{+,-,33\}) \ ,
\eal\ee
where
\be
v_{x3} = - \frac{\omega}{k} v_{t3} + \frac{\Lambda^2}{2kL^4} \left( s_+-s_- \right) \ .
\ee
Near the Poincar\'e horizon the ingoing-wave boundary conditions implies
\be\bal
\zeta_{3}&= e^{-L^2\frac{\sqrt{k^2-\omega^2}}{r}} r^{-1/2}  \left( \zeta_3^{(0)} + \cdots \right) \ , \\
\alpha_{t3}&= e^{-L^2\frac{\sqrt{k^2-\omega^2}}{r}} r^{-1/2}  \left( \alpha_{t3}^{(0)} + \cdots \right) \ , \\
\alpha_{x3}&= e^{-L^2\frac{\sqrt{k^2-\omega^2}}{r}} r^{-1/2}  \left( - \frac{\omega}{k} \alpha_{t3}^{(0)} + \cdots \right) \ , \\
\alpha_{J}&=  e^{-L^2\frac{\sqrt{k^2-\omega^2}}{r}} r^{-3/2}  \left( \alpha_J^{(0)} + \cdots \right)  \qquad\, (\text{with }J=\{+,-,33\}) \ ,
\eal\ee
where the five ${}^{(0)}$--superscripted symbols are the free coefficients that parameterise any IR-regular solution.

To find the quasi-normal modes we follow \cite{Amado:2009ts,Kaminski:2009dh}: we fix $\Lambda$ and $k$ to some values (we are using the rescaled variables of Eqn.~\eqref{eq.rescaling}) and we study the dependence on the complex frequency $\omega$. We build five independent ingoing wave solutions by shooting from the IR with 
\be
\{ \alpha_{t3}^{(0)}, \, \alpha_{+}^{(0)}, \, \alpha_{-}^{(0)}, \, \alpha_{33}^{(0)}, \, \zeta_3^{(0)} \} = \begin{cases}
\{ 1 , \, 1 , \, 1 , \, 1 , \, 1 \} \\
\{ 1 , \, -1 , \, 1 , \, 1 , \, 1 \} \\
\{ 1 , \, 1 , \, -1 , \, 1 , \, 1 \} \\
\{ 1 , \, 1 , \, 1 , \, -1 , \, 1 \} \\
\{ 1 , \, 1 , \, 1 , \, 1 , \, -1 \} 
\end{cases} \ .
\ee

Once the numeric solutions are obtained we can read the six source terms, $s_X$, in \eqref{eq.UVcoupled1}. The coefficients $s_3$, $s_{t3}$ and $s_{x3}$ are easy to get from the constant value the numeric solution tends to. To obtain the $s_J$, given the presence of the logarithm, it is better to focus instead on
\be
r^2 \left( r\, \alpha_J' + 2 \alpha_J \right) \simeq s_J + \cdots \ .
\ee
This allows to construct five 6-tuples, denoted here as $S_{1,\cdots,5}$, of asymptotic values for the sources. To determine the quasi-normal modes we want to find the values of the complex frequency such that a linear combinations of the $S_{1,\cdots,5}$ 6-tuples produces source-less ingoing solutions. Actually, since we are working with the gauge-dependent variables, we want that a linear combination of the $S_{1,\cdots,5}$ be proportional to the corresponding 6-tuple for the pure gauge solution \eqref{eq.gaugelambda3}. This means
\be
\det \begin{pmatrix} S_1 \\ S_2 \\ S_3 \\ S_4 \\ S_5 \\ S_{gauge} \end{pmatrix} = 0 \ .
\ee
Notice that these are two conditions (real and imaginary part of the determinant) for two variables (real and imaginary part of the frequency). Once we find a quasi-normal mode we can track how its position changes when $k$ or $\Lambda$ are varied.

\subsection{SO(2) scalars neutral under the diagonal of U(1)$_\Rsym \times$U(1)$_\iso$ and odd\label{app.gaugelambda12}}

This section describes how to calculate the quasi-normal modes in the channel coupling the following fluctuations: $\alpha_{t1}$, $\alpha_{t2}$, $\alpha_{x1}$, $\alpha_{x2}$, $\alpha_{13}$, $\alpha_{23}$, $\alpha_{31}$, $\alpha_{32}$, $\zeta_{1}$ and $\zeta_2$, subjected to two first-order constraints. Actually, it is a straightforward generalisation of the previous case, but in this case there are eight ingoing-wave solutions, eight outcoming-wave ones and two pure gauge solutions given by \eqref{eq.gaugelambda12} 

The UV asymptotics of the fields are
\be\bal\label{eq.UVcoupled2}
\zeta_{H}&=s_H + \frac{v_H}{r^2} + \cdots  \qquad \qquad\,\,\, (\text{with }H=\{1,2\}) \ , \\
\alpha_{I}&=s_I + \frac{v_I}{r^2} + \cdots  \qquad \qquad\quad (\text{with }I=\{t1,x1,t2,x2\}) \ , \\
\alpha_{J}&= \frac{s_I}{r^2} \log[r] + \frac{v_J}{r^2} + \cdots  \qquad\, (\text{with }J=\{13,23,31,32\}) \ ,
\eal\ee
where 
\be\bal
v_{x1} & = - \frac{\omega}{k} v_{t1} + \frac{i}{k} \left[ 2 \Mq \left( v_2 - v_{t2} \right) + 2\Mq \Lambda^2 \left( s_2 - s_{t2} \right) - \frac{\Lambda^2}{L^4} \left( s_{23} - s_{32} \right) \right] \ , \\
v_{x2} & = - \frac{\omega}{k} v_{t2} - \frac{i}{k} \left[ 2 \Mq \left( v_1 - v_{t1} \right) + 2 \Mq \Lambda^2 \left( s_1 - s_{t1} \right) - \frac{\Lambda^2}{L^4} \left( s_{13} - s_{31} \right) \right] \ .
\eal \ee
Near the Poincar\'e horizon the ingoing-wave boundary condition implies
\be\bal
\zeta_{H}&= e^{-L^2\frac{\sqrt{k^2-\omega^2}}{r}} r^{-1/2}  \left( \zeta_H^{(0)} + \cdots \right)\qquad\quad\, (\text{with }H=\{1,2\}) \ , \\
\alpha_{tH}&= e^{-L^2\frac{\sqrt{k^2-\omega^2}}{r}} r^{-1/2}  \left( \alpha_{tH}^{(0)} + \cdots \right)\qquad\quad (\text{with }H=\{1,2\}) \ , \\
\alpha_{xH}&= e^{-L^2\frac{\sqrt{k^2-\omega^2}}{r}} r^{-1/2}  \left( - \frac{\omega}{k} \alpha_{tH}^{(0)} + \cdots \right)\quad\,\, (\text{with }H=\{1,2\}) \ , \\
\alpha_{J}&=  e^{-L^2\frac{\sqrt{k^2-\omega^2}}{r}} r^{-3/2}  \left( \alpha_J^{(0)} + \cdots \right)  \qquad\quad (\text{with }J=\{13,23,31,32\}) \ ,
\eal\ee
where the eight ${}^{(0)}$--superscripted symbols are the free coefficients that parameterise any IR-regular solution. Quasi-normal modes are obtained exactly as before.


\end{document}